\documentclass[12pt]{article} 
\usepackage{amssymb}
\usepackage{graphicx}
\usepackage{fullpage}
\usepackage{subfigure}
\usepackage{algo}
\newtheorem{theorem}{Theorem}
\newtheorem{lemma}{Lemma}

\newtheorem{definition}{Definition}

\newcommand{\UDG}{\mathord{\it UDG}}
\newcommand{\PLDG}{\mathord{\it PLDG}}
\newcommand{\UDel}{\mathord{\it UDel}}
\newcommand{\Del}{\mathord{\it Del}}
\newcommand{\Int}{\mathord{\it int}}
\newcommand{\DT}{\mathord{\it DT}}
\newcommand{\LDT}{\mathord{\it LDT}}
\newcommand{\arc}{\mathord{\it arc}}
\newcommand{\aPLDG}{\textsf{\sc PLDG}}

\newcommand{\qed}{\rule{0.5em}{1.5ex}}
\newcommand{\fqed}{{\hfill~\qed}}
\newenvironment{proof}{{\noindent \bf Proof.}}
                      {{\hfill \fqed} \vspace{1em}}

\title{Communication-Efficient Construction of the Plane Localized 
Delaunay Graph}
\author{
Prosenjit Bose\thanks{School of Computer Science, Carleton University, 
        Ottawa, Canada. This work was supported by the Natural 
        Sciences and Engineering Research Council of Canada.} 
\and 
Paz Carmi\footnotemark[1]
\and 
Michiel Smid\footnotemark[1]
\and 
Daming Xu\footnotemark[1]
}
\date{\today}

\begin{document}

\maketitle

\begin{abstract}
Let $V$ be a finite set of points in the plane. We present a $2$-local 
algorithm that constructs a plane $\frac{4 \pi \sqrt{3}}{9}$-spanner of 
the unit-disk graph $\UDG(V)$. This algorithm makes only one round of  
communication and each point of $V$ broadcasts at most $5$ messages. 
This improves the previously best message-bound of $11$ by 
Ara\'{u}jo and Rodrigues (Fast localized Delaunay triangulation, 
Lecture Notes in Computer Science, volume 3544, 2004). 
\end{abstract}

%%%%%%%%%%%%%%%%%%%%%%%%%%%%%%%%%%%%%%%%%%%%%%%%%%%%%%%%%%%%%%%%%%%%%%%%
\section{Introduction} \label{INTRO}
%%%%%%%%%%%%%%%%%%%%%%%%%%%%%%%%%%%%%%%%%%%%%%%%%%%%%%%%%%%%%%%%%%%%%%%%

A \emph{wireless ad hoc network} consists of a finite set $V$ of wireless 
nodes. Each node $u$ in $V$ is a point in the plane that can communicate 
directly with all points of $V$ within $u$'s communication range. 
If this range is one unit for each point, then the network is modeled 
by the \emph{unit-disk graph} $\UDG(V)$ of $V$. This (undirected) graph 
has $V$ as its vertex set and any two distinct vertices $u$ and $v$ 
are connected by an edge if and only if the Euclidean distance 
$|uv|$ between $u$ and $v$ is at most one unit.    

In order for two points that are more than one unit apart to be able to 
communicate, the points of $V$ use a so-called \emph{local algorithm}  
(to be defined below) to construct a subgraph $G$ of $\UDG(V)$. This 
subgraph should have the property that it supports efficient routing of 
messages, i.e., there should be a simple and efficient protocol that 
allows any point of $V$ to send a message to any other point of $V$. 

In this paper, we present a local algorithm that constructs a subgraph 
$G$ of $\UDG(V)$ that satisfies the following properties: 
\begin{enumerate} 
\item Each point $u$ of $V$ stores a set $E(u)$ of edges that are 
      incident on $u$. The edge set of $G$ is equal to 
      $\cup_{u \in V} E(u)$. 
\item The edge sets $E(u)$ with $u \in V$ are \emph{consistent}: 
      For any two points $u$ and $v$ in $V$, $(u,v)$ is an edge in 
      $E(u)$ if and only if $(u,v)$ is an edge in $E(v)$.    
\item The graph $G$ is \emph{plane}: If we consider each edge 
      $(u,v)$ to be the straight-line segment joining $u$ and $v$, 
      then no two edges of $G$ 
      cross\footnote{Two edges are said to \emph{cross} if they are not 
      collinear and there exists a (unique) point that is in the 
      relative interior of both edges.}. The graph 
      being plane is useful, because several algorithms are known for 
      routing messages in a plane subgraph of $\UDG(V)$; see, e.g., 
      Bose \emph{et al.}~\cite{bmsu-wn-01} and 
      Karp and Kung~\cite{kk-mobicom-00}.  
\item The graph $G$ is a $t$-\emph{spanner} of $\UDG(V)$, for some 
      constant $t>1$: For each edge $(u,v)$ of $\UDG(V)$, the graph 
      $G$ contains a path between $u$ and $v$ whose Euclidean length 
      is at most $t|uv|$. Observe that this implies that shortest-path 
      distances in $\UDG(V)$ are approximated, within a factor of $t$, 
      by shortest-path distances in $G$. Thus, this property implies 
      that the total distance traveled by a message, when using $G$, 
      is not much larger than the minimum distance that needs to be 
      traveled in $\UDG(V)$. 
\end{enumerate}

%%%%%%%%%%%%%%%%%%%%%%%%%%%%%%%%%%%%%%%%%%%%
\subsection{Local Algorithms}\label{MODEL}
%%%%%%%%%%%%%%%%%%%%%%%%%%%%%%%%%%%%%%%%%%%%

As mentioned above, we model a wireless ad hoc network by the unit-disk 
graph $\UDG(V)$, where $V$ is a finite set of points in the plane. 
The points of $V$ want to construct a communication graph $G$ (which is 
a subgraph of $\UDG(V)$) using a distributed and local algorithm. 
In this section, we formalize this notion and introduce the 
complexity measures that we will use to analyze the efficiency of 
such algorithms.    

The points of $V$ can communicate with each other by broadcasting 
messages. If a point $u$ of $V$ broadcasts a message, then each point of 
$V$ within Euclidean distance one from $u$ receives the message. Each 
point of $V$ can perform computations based on its location and all 
information received from other points. Informally, an algorithm is 
called \emph{local}, if the computation performed at each point $u$ of 
$V$ is based only on its location and the locations of all points that 
are within distance $k$ (in $\UDG(V)$) from $u$, for some small integer 
$k \geq 1$. Thus, in a local algorithm, information cannot ``travel'' 
over a ``large'' distance.

To define this notion formally, let $\delta_{\UDG}(u,v)$ denote the 
Euclidean length of a shortest path between the points $u$ and $v$ in 
the graph $\UDG(V)$. For any integer $k \geq 1$, let  
\[ N_k(u) = \{ v \in V : \delta_{\UDG}(u,v) \leq k \} . 
\]
Observe that $u \in N_k(u)$. 

Let $\mathcal{A}(V)$ be a distributed algorithm that runs on a set $V$ 
of points in the plane, and let $\mathcal{A}(u;V)$ denote the computation 
performed by point $u$. As is common in this field, we assume that, at 
the start of the algorithm, each point $u$ of $V$ knows the locations 
(i.e., the $x$- and $y$-coordinates) of all points in $N_1(u)$. Thus, 
the set $N_1(u)$ can be considered to be the input for $u$. 

For any point $u$ of $V$, we denote by $T_u(V)$ the \emph{trace} of the 
computation performed by $\mathcal{A}(u;V)$. Thus, $T_u(V)$ contains the 
sequence of all computing and broadcasting operations performed by 
$\mathcal{A}(u;V)$ when each point $v$ of $V$ runs algorithm 
$\mathcal{A}(v;V)$. 

\begin{definition} 
For an integer $k \geq 1$, we say that $\mathcal{A}(V)$ is a 
$k$-\emph{local} algorithm, if for each point $u$ of $V$, 
\[ T_u(V) = T_u( N_k(u) ) .  
\] 
\end{definition} 

In other words, for every point $u$ of $V$, the following holds: If we 
run the \emph{entire} distributed algorithm $\mathcal{A}$ with $V$ 
replaced by $N_k(u)$, then the computation performed by $u$ does not 
change (even though the computations performed by other points may 
change). 

A $k$-local algorithm runs in parallel on all points of $V$, where each 
point $u$ performs an alternating sequence of computation steps and 
broadcasting steps in a synchronized manner. In a 
\emph{computation step}, point $u$ performs some computation based on 
the subset of $N_k(u)$ that is known to $u$ at that moment. 
(For example, $u$ may compute the Delaunay triangulation of this 
subset; we consider this to be one computation step.) In a 
\emph{broadcasting step}, point $u$ broadcasts a (possibly empty) 
sequence of messages, which is received by all points in $N_1(u)$. 
In this paper, a \emph{message} is defined to be the location of a 
point in the plane (which need not be an element of $V$). 
The efficiency of a local algorithm will be expressed in terms of the 
following measures:   

\begin{enumerate} 
\item The value of $k$. The smaller the value of $k$, the ``more local'' 
      the algorithm is. 
\item The maximum number of messages that are broadcast by any point 
      of $V$. The goal is to minimize this number.  
\item The number of \emph{communication rounds}, which is defined to be 
      the maximum number of broadcasting steps performed by any point in 
      $V$. This number measures the (parallel) time for the entire 
      algorithm to complete its computation. Again, the goal is to 
      minimize this number.  
\end{enumerate}

%%%%%%%%%%%%%%%%%%%%%%%%%%%%%%%%%%%%%%%
\subsection{Previous Work}\label{PREWK}
%%%%%%%%%%%%%%%%%%%%%%%%%%%%%%%%%%%%%%%

Above, we have defined the notion of a $t$-spanner of the unit-disk 
graph $\UDG(V)$. For a real number $t>1$, a graph $G$ is called a 
$t$-spanner of the \emph{point set} $V$ if for any two elements $u$ 
and $v$ of $V$, there exists a path in $G$ between $u$ and $v$ whose 
length is at most $t|uv|$. The problem of constructing $t$-spanners 
for point sets has been studied intensively in computational geometry; 
see the book by Narasimhan and Smid~\cite{ns-gsn-07} for a survey. 

Since we are concerned with plane spanners of the unit-disk graph, our 
algorithm will be based on the \emph{Delaunay Triangulation} $\DT(V)$ of 
$V$; see, e.g., the textbook by de 
Berg \emph{et al.}~\cite{bcko-cgaa-08}. Recall that $\DT(V)$ is the 
plane graph with vertex set $V$ in which any two distinct points $u$ 
and $v$ are connected by an edge if and only if there exists a disk 
$D$ such that (i) $u$ and $v$ are the only points of $V$ that are on 
the boundary of $D$ and (ii) no point of $V$ is in the interior of $D$. 
Also, three points $u$, $v$, and $w$ determine a triangular face of 
$\DT(V)$ if and only if the disk having $u$, $v$, and $w$ on its 
boundary does not contain any point of $V$ in its interior. 
Keil and Gutwin~\cite{kg-cgwac-92} have shown that $\DT(V)$ is a 
$\frac{4 \pi \sqrt{3}}{9}$-spanner of $V$. To extend this result to 
unit-disk graphs, it is natural to consider subgraphs of $\UDel(V)$, 
which is defined to be the intersection of the Delaunay triangulation 
and the unit-disk graph of $V$. It has been shown by 
Bose \emph{et al.}~\cite{bmnsz-agbsp-03} that $\UDel(V)$ is a 
$\frac{4 \pi \sqrt{3}}{9}$-spanner of $\UDG(V)$. Unfortunately, 
constructing $\UDel(V)$ using a $k$-local algorithm, for any constant 
value of $k$, is not possible: Consider an edge $(u,v)$ in $\UDel(V)$ 
whose empty disk $D$ is very large. In order for a $k$-local algorithm 
to verify that no point of $V$ is in the interior of $D$, information 
about the points of $V$ must travel over a large distance to $u$ or 
$v$. Clearly, this is possible only if the value of $k$ is very 
large. Because of this, researchers have considered the problem of 
designing local algorithms that construct a plane subgraph of $\UDG(V)$ 
which is a \emph{supergraph} of $\UDel(V)$. Obviously, by the result 
of~\cite{bmnsz-agbsp-03}, such a graph is also a 
$\frac{4 \pi \sqrt{3}}{9}$-spanner of $\UDG(V)$. 

Gao \emph{et al.}~\cite{gghzz-ieeejsac-05} proposed a $2$-local 
algorithm that constructs a plane subgraph of $\UDG(V)$ which is a 
supergraph of $\UDel(V)$. However, the number of messages broadcast by 
a single point of $V$ can be as large as $\Theta(n)$, where $n$ is the 
number of elements of $V$. This result was improved by 
Li \emph{et al.}~\cite{lcww-ieeepds-03}: They presented a $2$-local 
algorithm that constructs such a graph in four communication rounds 
and in which each point broadcasts at most $49$ messages. 

Currently, the best result for computing a plane $t$-spanner (for some 
constant $t$) of the unit-disk graph $\UDG(V)$ is by 
Ara\'{u}jo \emph{et al.}~\cite{ar-opodis-05}. They presented a $2$-local 
algorithm which computes such a spanner in one communication round and 
in which each point broadcasts at most $11$ messages.

%%%%%%%%%%%%%%%%%%%%%%%%%%%%%%%%%%%%%%%%%%%%%
\subsection{Our Result}  \label{ourresults}
%%%%%%%%%%%%%%%%%%%%%%%%%%%%%%%%%%%%%%%%%%%%%

In this paper, we improve the upper bound of 
Ara\'{u}jo \emph{et al.}~\cite{ar-opodis-05} on the message complexity 
for each point of $V$ from $11$ to $5$:  

\begin{theorem}   \label{thm1}
       Let $V$ be a finite set of points in the plane. There exists a 
       $2$-local algorithm that computes a plane and consistent  
       $\frac{4 \pi \sqrt{3}}{9}$-spanner of the unit-disk graph of $V$. 
       This algorithm makes one communication round and each point of 
       $V$ broadcasts at most $5$ messages.  
\end{theorem}  

The rest of this paper is organized as follows. In Section~\ref{PLDEL},  
we present a preliminary $2$-local algorithm that computes, in one 
communication round, a subgraph of $\UDG(V)$. In this algorithm, each 
point of $V$ broadcasts at most $6$ messages. We present a 
\emph{rigorous} proof of the fact that the graph computed by this 
algorithm is a plane and consistent $\frac{4 \pi \sqrt{3}}{9}$-spanner 
of $\UDG(V)$. In Section~\ref{improve}, we make a simple modification 
to the algorithm of Section~\ref{PLDEL} which reduces the message 
complexity for each point of $V$ from $6$ to $5$. We then show that the 
new algorithm and the algorithm of Section~\ref{PLDEL} compute the same 
graph. Thus, this will prove Theorem~\ref{thm1}. We conclude in 
Section~\ref{CONCL} with some directions for future work. 

Throughout the rest of this paper, we assume that the points in the 
set $V$ are in general position (meaning that no three points of $V$ 
are collinear and no four points of $V$ are cocircular). We also 
assume that the unit-disk graph $\UDG(V)$ is connected.  
We will use the following notation: 
\begin{itemize} 
\item $D(a,b,c)$ denotes the disk having the three points $a$, $b$, and 
      $c$ on its boundary. 
\item $D(c;r)$ denotes the disk centered at the point $c$ and having 
      radius $r$. 
\item $\Delta(a,b,c)$ denotes the triangle having the three points $a$, 
      $b$, and $c$ as its vertices.
\item $\partial D$ denotes the boundary of the disk $D$. 
\item $\Int(D)$ denotes the interior of the disk $D$. 
\item Let $v$, $x$, and $y$ be points of $V$, where $v \neq y$. Assume 
      there exists a disk $D$ such that 
      $N_1(x) \cap \partial D = \{v,y\}$ and 
      $N_1(x) \cap \Int(D) = \emptyset$. 
      We denote such a disk $D$ by $\Del_x(v,y)$.  
      Observe that $\Del_x(v,y)$ is a certificate for the fact that 
      $(v,y)$ is an edge in the Delaunay triangulation of the point 
      set $N_1(x)$. 
\end{itemize}

%%%%%%%%%%%%%%%%%%%%%%%%%%%%%%%%%%%%%%%%%%%%%%%%%%%%%%%%%%%%%%%%%%%%%%%%
\section{A Preliminary Algorithm}
\label{PLDEL}
%%%%%%%%%%%%%%%%%%%%%%%%%%%%%%%%%%%%%%%%%%%%%%%%%%%%%%%%%%%%%%%%%%%%%%%%

In this section, we present a $2$-local algorithm that constructs a 
graph, called the \emph{plane localized Delaunay graph} $\PLDG(V)$, 
whose vertex set is a finite set $V$ of points in the plane. The 
algorithm computes $\PLDG(V)$ in one communication round and each point 
of $V$ broadcasts at most $6$ messages. We will prove that $\PLDG(V)$ 
is a plane and consistent supergraph of $\UDel(V)$. 

In the construction, each point $v$ of $V$ runs algorithm $\aPLDG(v)$ 
in parallel. Let $N_v = N_1(v)$, i.e., 
$N_v = \{ u \in V : |uv| \leq 1\}$. Recall that we assume that, at the 
start of the algorithm, point $v$ knows the locations of all points in 
$N_v$. Algorithm $\aPLDG(v)$ first computes the Delaunay triangulation 
$\LDT(v)$ of the set $N_v$. Then, for each triangular face 
$\Delta(u,v,w)$ in $\LDT(v)$ for which $\angle{uvw} > \frac{\pi}{3}$, 
algorithm $\aPLDG(v)$ broadcasts the location $v$ together with the 
center of the disk $D(u,v,w)$ containing $u$, $v$, and $w$ on its 
boundary. 

In the final step, algorithm $\aPLDG(v)$ checks the validity of all 
edges that are incident on $v$ in $\LDT(v)$ and removes those edges 
which cause a crossing. To be more precise, let $x$ be a point in $N_v$, 
and assume that $v$ receives a center $c'_i$ from $x$. 
Algorithm $\aPLDG(v)$ considers the unit-disk $D(v;1)$ centered at $v$
and the disk $D(c'_i; |c'_ix|)$ centered at $c'_i$ that contains $x$ on 
its boundary. The algorithm knows that $\partial D(c'_i; |c'_ix|)$ 
contains exactly three points which define a triangular face in the 
Delaunay triangulation $\LDT(x)$ of $N_x$. Point $x$ is one of these 
three points; let $p$ and $q$ be the other two points. Assume that the 
set $N_v$ contains exactly two points of $\{x,p,q\}$, say $x$ and $p$. 
Thus, algorithm $\aPLDG(v)$ knows the points $x$ and $p$, but it does 
not know $q$. The algorithm computes $\arc_i$, which is defined to be 
the (open) portion of $\partial D(c'_i; |c'_ix|)$ which is not contained 
in $D(v;1)$. Even though the algorithm does not know the exact location 
of the third point $q$, it does know that $q$ is on $\arc_i$. The 
algorithm chooses an arbitrary point $z'$ on $\arc_i$ such that 
$|x z'| \leq 1$ or $|p z'| \leq 1$ and \emph{acts as if} 
$\Delta(x,p,z')$ is a triangular face in $\LDT(x)$. (Observe that, 
since $q \in \arc_i$ and $|xq| \leq 1$, the algorithm can choose such 
a point $z'$. Also, $z'$ is not necessarily a point of $V$.) The 
algorithm now considers each edge $(v,y)$ in $\LDT(v)$ (where, possibly, 
$v=p$, $y=p$, or $y=x$) and uses the triangle $\Delta(x,p,z')$ to 
decide whether or not to remove $(v,y)$: Since $(v,y)$ is an edge in 
$\LDT(v)$, algorithm $\aPLDG(v)$ can compute a disk $D = \Del_v(v,y)$ 
such that (i) $v$ and $y$ are the only points of $N_v$ that are on the 
boundary of $D$ and (ii) the interior of $D$ does not contain any point 
of $N_v$. If $\arc_i$ is fully contained in the interior of 
$\Del_v(v,y)$, then the algorithm knows that $q$ is contained in the 
interior of $\Del_v(v,y)$ (even though it does not know the exact 
location of $q$) and, therefore, $\Del_v(v,y)$ is not a certificate 
that $(v,y)$ is an edge in the Delaunay triangulation of the entire 
set $V$. The algorithm now checks if the line segment $vy$ crosses 
any of the two line segments $xz'$ and $pz'$ and, if so, removes the 
edge $(v,y)$. Observe that if $(v,y)$ is not an edge of the Delaunay 
triangulation $\DT(V)$, the algorithm still keeps it as long as it 
does not cross any other edge.  

\begin{figure}
\begin{algorithm}{\sc PLDG}[v]{\label{algo1}}
let $N_v = \{ u \in V : |uv| \leq 1\}$;\\
compute the Delaunay triangulation $\LDT(v)$ of $N_v$;\\
let $E(v)$ be the set of all edges in $\LDT(v)$ that are incident 
on $v$;\\
let $\Delta_v$ be the set of all triangular faces $\Delta(u,v,w)$ in 
$\LDT(v)$ for which $\angle{uvw} > \frac{\pi}{3}$;\\ 
let $k$ be the number of elements in $\Delta_v$;\\ 
\qif $k \geq 1$\\
\qthen let $c_1,\ldots,c_k$ be the centers of the circumcircles of all 
       triangles in $\Delta_v$;\\
       broadcast the sequence $(v, c_1, \ldots, c_k)$;
\qendif\\
\qfor each sequence $(x, c'_1, \ldots, c'_m)$ received\\
\qdo \qfor $i=1$ \qto $m$\\
     \qdo let $D(c'_i;|c'_ix|)$ be the disk with center $c'_i$ that 
          contains $x$ on its boundary;\\
          \qif $\partial D(c'_i;|c'_ix|)$ contains exactly two points 
               of $N_v$ \\
          \qthen let $p$ be the point in 
                 $(N_v \setminus \{x\}) \cap \partial D(c'_i;|c'_ix|)$;\\  
                 let $\arc_i$ be the (open) arc on 
                 $\partial D(c'_i;|c'_ix|)$ that is not contained in 
                 the unit-disk $D(v;1)$ centered at $v$;\\
                 let $z'$ be an arbitrary point on $\arc_i$ with 
                 $|x z'| \leq 1$ or  $|p z'| \leq 1$;\\ 
                 \qfor each edge $(v,y)$ in $E(v)$\\ 
                 \qdo let $\Del_v(v,y)$ be a disk $D$ such that 
                      $N_v \cap \partial D = \{v,y\}$ and 
                      $N_v \cap \Int(D) = \emptyset$; \\ 
                      \qif $\arc_i$ is contained in the interior of 
                           $\Del_v(v,y)$ and the line segment $vy$ 
                           crosses at least one of the line segments 
                           $xz'$ and $pz'$ \\
                      \qthen remove $(v,y)$ from $E(v)$ 
                      \qendif
                 \qendfor
          \qendif
     \qendfor
\qendfor 
\end{algorithm}
\caption{\sl The plane localized Delaunay graph algorithm.}
\label{figPLDEL}
\end{figure}

\begin{figure}
  \subfigure[edge $(v,y)$ is removed, where $y \neq x$, $y \neq p$, 
             and $v \neq p$.]{
    \label{fig02:a}
    \begin{minipage}[b]{0.5\textwidth}
      \centering
      \includegraphics[scale=0.37]{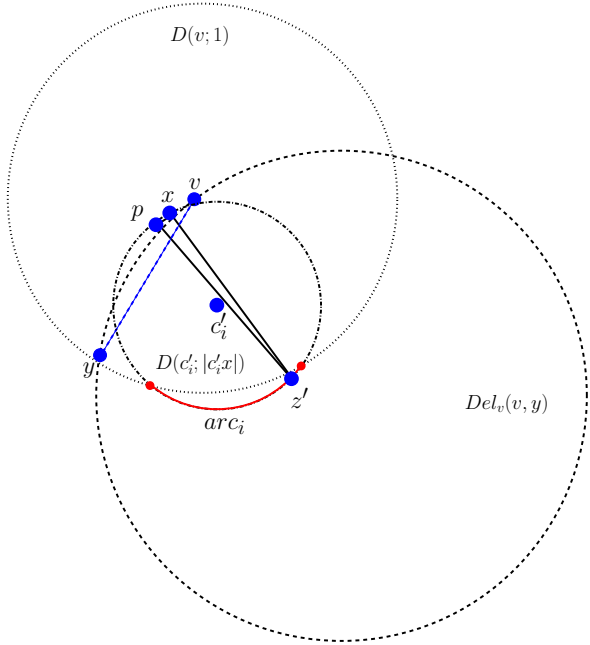}
    \end{minipage}}%
  \subfigure[edge $(v,y)$ is removed, where $v=p$.]{
    \label{fig02:b}
    \begin{minipage}[b]{0.5\textwidth}
      \centering
      \includegraphics[scale=0.37]{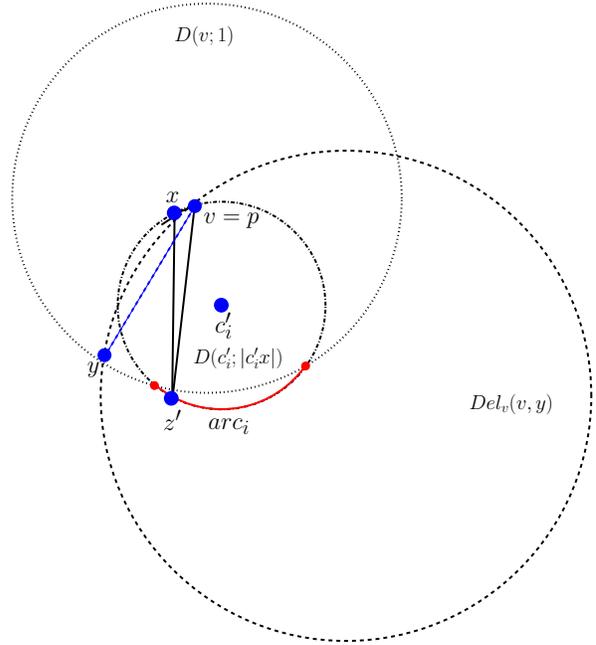}
    \end{minipage}}%

  \subfigure[edge $(v,y)$ is removed, where $y=p$.]{
    \label{fig02:c}
    \begin{minipage}[b]{0.5\textwidth}
      \centering
      \includegraphics[scale=0.37]{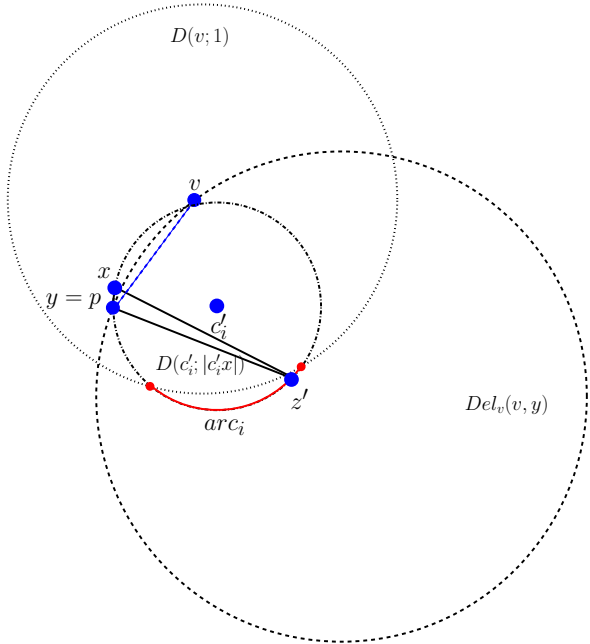}
    \end{minipage}}%
  \subfigure[edge $(v,y)$ is removed, where $y=x$.]{
    \label{fig02:d}
    \begin{minipage}[b]{0.5\textwidth}
      \centering
      \includegraphics[scale=0.37]{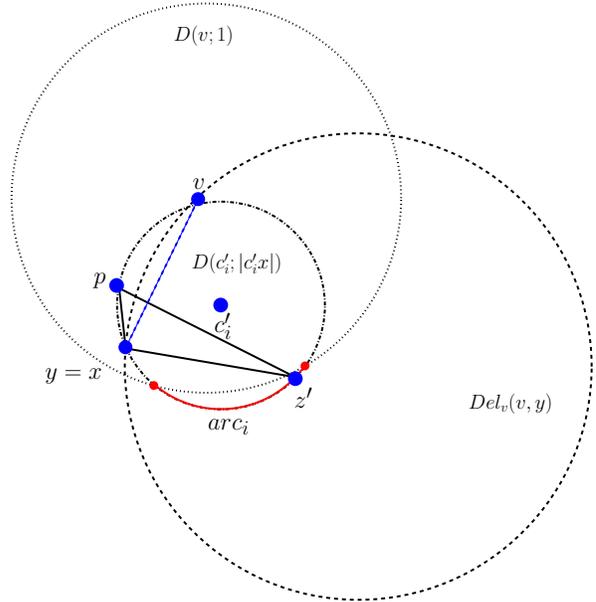}
    \end{minipage}}%
  \caption{\sl Illustrating algorithm $\aPLDG(v)$.}
  \label{fig02}
\end{figure}

The formal algorithm is given in Figure~\ref{figPLDEL}. An illustration,  
with the special cases when $v=p$, $y=p$, or $y=x$, is given in 
Figure~\ref{fig02}. 

Running algorithm $\aPLDG(v)$ for all points $v$ of $V$ in parallel will 
be referred to as running algorithm $\aPLDG(V)$. 
We denote by $E(v)$ the edge set that is computed by algorithm 
$\aPLDG(v)$. Observe that each edge in $E(v)$ is incident on the point  
$v$. Let $E = \cup_{v \in V} E(v)$ and let $\PLDG(V)$ denote the graph 
with vertex set $V$ and edge set $E$. 

In the rest of this section, we will prove a sequence of lemmas which 
lead to the proof that $\PLDG(V)$ is a plane and consistent supergraph 
of $\UDel(V)$; see Lemmas~\ref{lem05}, \ref{lem07} and~\ref{lem08}. 

\begin{figure}
  \centering
  \includegraphics[scale=0.55]{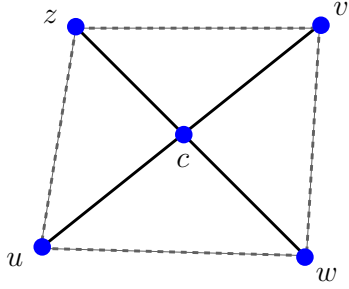}
  \caption{\sl An illustration of Lemma~\ref{lem01}. At least one of 
       $\{u, v, w, z\}$ is within distance $1$ from all other points.}
  \label{fig01}
\end{figure}

We start with a simple, but fundamental lemma:  

\begin{lemma}   \label{lem01}
Let $S=\{u,v,w,z\}$ be a set of four points in the plane in general 
position, such that $|uv| \leq 1$, $|wz| \leq 1$, and the line segments 
$uv$ and $wz$ cross. Then there exists a point $x$ in $S$ such that 
$|xy| \leq 1$ for all $y$ in $S$. 
\end{lemma} 
\begin{proof}
Let $c$ be the intersection of the line segments $uv$ and $wz$; see 
Figure~\ref{fig01}. By the triangle inequality, we have 
$|uw| \leq |uc|+|cw|$ and $|vz| \leq |vc|+|cz|$. Since 
$|uv|=|uc|+|cv| \leq 1$ and $|wz|=|wc|+|cz| \leq 1$, we have 
$|uw|+|vz| \leq |uv|+|wz|\leq 2$. Therefore, at least one of $uw$ and 
$vz$ has length at most $1$. Without loss of generality, assume that 
$|uw| \leq 1$. By a symmetric argument, at least one of $uz$ and $vw$ 
has length at most $1$. Without loss of generality, assume that 
$|uz| \leq 1$. Then all distances $|uv|$, $|uw|$, and $|uz|$ are at 
most $1$ and, thus, we can take $x$ to be the point $u$.  
\end{proof}

The following lemma implies that for every edge $(v,y)$ in $E(v)$, 
the edge $(v,y)$ is in the Delaunay triangulation $\LDT(y)$ of the 
set $N_y$. 

\begin{lemma}\label{lem02} 
       Let $v$ and $y$ be two distinct points of $V$ and assume that 
       $(v,y)$ is not an edge in $\LDT(y)$. Then, after algorithm 
       $\aPLDG(V)$ has terminated, $(v,y)$ is not an edge in $E(v)$. 
\end{lemma} 
\begin{proof}
First assume that $(v,y)$ is not an edge in $\LDT(v)$. Then, since $E(v)$ 
is a subset of the edge set of $\LDT(v)$, $(v,y)$ is not an edge in 
$E(v)$. 

\begin{figure}
  \centering
  \includegraphics[scale=0.60]{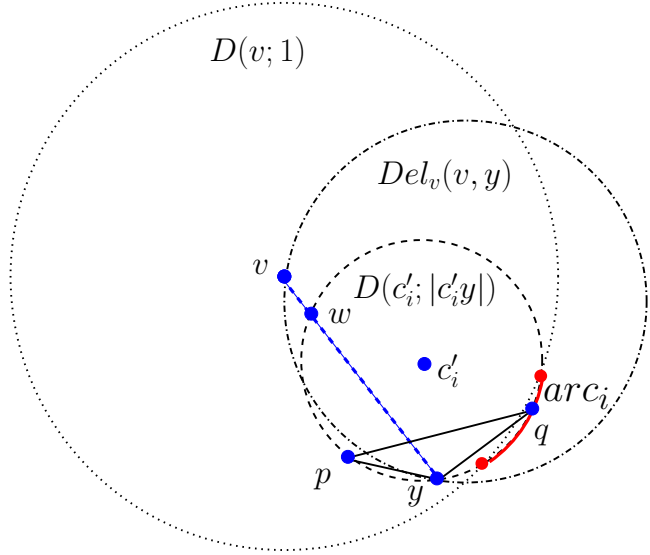}
  \caption{\sl An illustration of the proof of Lemma \ref{lem02}.}  
  \label{fig03}
\end{figure}

From now on, we assume that $(v,y)$ is an edge in $\LDT(v)$. Observe 
that $|vy| \leq 1$. Since $(v,y)$ is not an edge in $\LDT(y)$, there 
exist two points $p$ and $q$ in $V$ such that the triangle 
$\Delta(y,p,q)$ is a triangular face in $\LDT(y)$ and $vy$ crosses 
$pq$; see Figure~\ref{fig03}. Observe that the points $p$, $q$, $v$, 
and $y$ are pairwise distinct. In the rest of the proof, we will do the 
following: 
\begin{enumerate} 
\item We first show that algorithm $\aPLDG(y)$ broadcasts the center of 
      the circumcircle of $\Delta(y,p,q)$. Since $|vy| \leq 1$, $v$ will 
      receive this center.    
\item We then show that, when algorithm $\aPLDG(v)$ considers the center 
      of $\Delta(y,p,q)$, it deletes the edge $(v,y)$. As a result, the 
      edge $(v,y)$ is not in $E(v)$. 
\end{enumerate}   

Let $c'_i$ be the center of the circumcircle of $\Delta(y,p,q)$ and 
consider the corresponding disk $D(c'_i;|c'_iy|)$, i.e., the disk with 
center $c'_i$ that contains $y$, $p$, and $q$ on its boundary. 
Recall that $D(v;1)$ denotes the unit-disk centered at $v$ and 
$N_v = \{ u \in V : |uv| \leq 1\}$. 

Since $|vy| \leq 1$ and $\Delta(y,p,q)$ is a triangular face in 
$\LDT(y)$, $v$ is not contained in $D(c'_i;|c'_iy|)$. Since $vy$ crosses 
$pq$, this implies that any disk $D$ with $v$ and $y$ on its boundary 
contains at least one of $p$ and $q$ (otherwise, $\partial D$ and 
$\partial D(c'_i;|c'_iy|)$ intersect more than twice). 

We first claim that $\partial D(c'_i;|c'_iy|)$ contains exactly two 
points of $N_v$. Since $y \in N_v$, this means that we claim that 
exactly one of $p$ and $q$ is in $N_v$. We prove this by contradiction. 
First assume that neither $p$ nor $q$ is in $N_v$. Then both $p$ and 
$q$ are outside $D(v;1)$. Let $D$ be the disk with diameter $vy$. 
Since $|vy| \leq 1$, $D$ is contained in $D(v;1)$. Thus, neither $p$ 
nor $q$ is contained in $D$, which is a contradiction, because $D$ 
contains $v$ and $y$ on its boundary. Now assume that both $p$ and $q$ 
are in $N_v$. Then, since any disk with $v$ and $y$ on its boundary 
contains one of $p$ and $q$, it follows that $(v,y)$ is not an edge 
in $\LDT(v)$, which is again a contradiction. 

Thus, we have shown that $\partial D(c'_i;|c'_iy|)$ contains exactly 
two points of $N_v$. We may assume without loss of generality that 
$y,p \in N_v$ and $q \not\in N_v$. 

Consider the triangle $\Delta(v,y,q)$. Since $|yv| \leq 1$, 
$|yq| \leq 1$, and $|vq|  > 1$, we have $\angle{vyq}>\frac{\pi}{3}$. 
Since $\angle{pyq} > \angle{vyq}$, it follows that 
$\angle{pyq} > \frac{\pi}{3}$. Since $\Delta(y,p,q)$ is a triangular 
face in $LDT(y)$, algorithm $\aPLDG(y)$ broadcasts a sequence in line~8 
which contains the center $c'_i$ of $D(c'_i;|c'_iy|)$. 

As we have mentioned above, since $|vy| \leq 1$, $v$ receives the 
sequence broadcast by $\aPLDG(y)$. This sequence contains the center 
$c'_i$ together with the point $y$. When algorithm $\aPLDG(v)$ considers 
$c'_i$, it discovers that $\partial D(c'_i;|c'_iy|)$ contains exactly 
two points of $N_v$; as we have seen above, these points are $y$ and 
$p$. Thus, the condition in line~12 is satisfied. In line~14, algorithm 
$\aPLDG(v)$ computes the open arc $\arc_i$, which is the part of 
$\partial D(c'_i;|c'_iy|)$ that is not contained in $D(v;1)$. Observe 
that even though $\aPLDG(v)$ does not know the location of the point 
$q$, the algorithm knows that it is on $\arc_i$. Let $\Del_v(v,y)$ be 
the disk that is computed by $\aPLDG(v)$ in line~17. This disk has the 
properties that $N_v \cap \partial \Del_v(v,y) = \{v,y\}$ and 
$N_v \cap \Int(\Del_v(v,y)) = \emptyset$. 

We show that $\arc_i$ is contained in the interior of $Del_v(v,y)$; 
thus, the first condition in line~18 is satisfied. Let $w$ be the 
intersection between $vy$ and $\partial D(c'_i;|c'_iy|)$, let 
$\widehat{wpy}$ be the arc on $\partial D(c'_i;|c'_iy|)$ with endpoints 
$w$ and $y$ and which contains $p$, and let $\widehat{yqw}$ be the arc 
on $\partial D(c'_i;|c'_iy|)$ with endpoints $y$ and $w$ and which 
contains $q$. Since $|vy| \leq 1$, $|vq|>1$, and $q \in \widehat{yqw}$, 
we have $\widehat{wpy} \subseteq D(v;1)$ (because otherwise, 
$\partial D(v;1)$ and $\partial D(c'_i;|c'_iy|)$ intersect more than 
twice). It follows that $arc_i \subseteq \widehat{yqw}$. Since 
$|vp| \leq 1$, we have $p \not\in \Del_v(v,y)$. Therefore, 
$\partial \Del_v(v,y)$ and $\widehat{wpy}$ intersect twice. Since 
$\partial \Del_v(v,y)$ and $\partial D(c'_i;|c'_iy|)$ cannot intersect 
more than twice, it follows that $\arc_i$ is contained in the interior 
of $\Del_v(v,y)$. 

Consider the point $z'$ on $\arc_i$ that is chosen in line~15 of 
algorithm $\aPLDG(v)$. We will show that $vy$ crosses $pz'$; thus, the 
second condition in line~18 is also satisfied. Assume, by contradiction, 
that $vy$ does not cross $pz'$. Since the line through $v$ and $y$ 
separates $p$ from the two points $q$ and $z'$, and since $vy$ crosses 
$pq$, it follows that $y$ or $v$ is in the triangle $\Delta(p,q,z')$. 
However, since $y$, $p$, $q$, and $z'$ are on the circle 
$\partial D(c'_i;|c'_iy|)$, $y$ cannot be in $\Delta(p,q,z')$. Also, 
since $\Delta(p,q,z')$ is contained in $D(c'_i;|c'_iy|)$ and since 
$v \in N_y$, $v$ cannot be in $\Delta(p,q,z')$, because otherwise, 
$\Delta(y,p,q)$ would not be a triangular face in $\LDT(y)$. Thus, we 
have shown that $vy$ crosses $pz'$. 

By inspecting algorithm $\aPLDG(v)$, it follows that it removes, in 
line~19, the edge $(v,y)$ from the edge set $E(v)$. This completes the 
proof. 
\end{proof}

\begin{figure}
  \centering
  \includegraphics[scale=0.55]{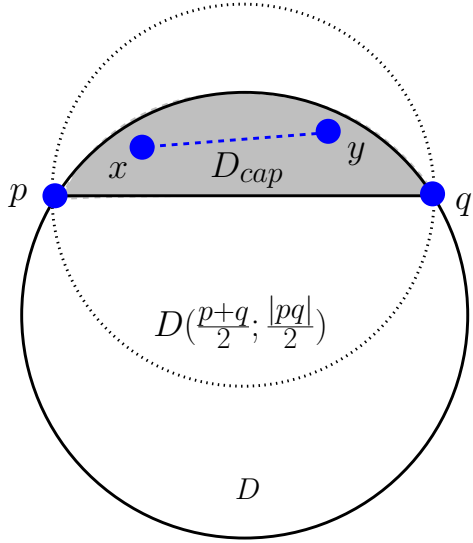}
  \caption{\sl An illustration of the proof of Lemma \ref{lem03}.}  
  \label{fig05}
\end{figure}

For the following geometric lemma, refer to Figure~\ref{fig05}. 

\begin{lemma}\label{lem03}
       Let $p$ and $q$ be two points with $|pq| \leq 1$, let $D$ be a 
       disk containing $p$ and $q$ on its boundary, and let $D_{cap}$ 
       be the part of $D$ that is bounded by the line segment $pq$ and 
       the minor arc $\widehat{pq}$ on $\partial D$ between $p$ and $q$. 
       Then $|xy| \leq 1$ for all $x$ and $y$ in $D_{cap}$.
\end{lemma} 
\begin{proof}
Consider the disk $D' = D(\frac{p+q}{2};\frac{|pq|}{2})$ with diameter 
$pq$. Since $\widehat{pq}$ is the minor arc on $\partial D$ between $p$ 
and $q$, $D_{cap}$ is completely contained in $D'$. Therefore, if $x$ 
and $y$ are points in $D_{cap}$, then these points are contained in 
$D'$. Since the diameter of $D'$ is at most $1$, it follows that 
$|xy| \leq 1$.
\end{proof}

The next lemma will form the basis for our claim that the graph 
$\PLDG(V)$ is plane. 

\begin{lemma}\label{lem04}
       Let $x$, $q$, $v$, and $y$ be four pairwise distinct points of 
       $V$. Assume that $|xq| \leq 1$, $|xv| \leq 1$, $|xy|\leq 1$, 
       $|vy| \leq 1$, $xq$ crosses $vy$, $(x,q)$ is an edge in 
       $\LDT(x)$, and $(v,y)$ is an edge in $\LDT(y)$. 
       Then, after algorithm $\aPLDG(V)$ has terminated, $(v,y)$ is 
       not an edge in $E(y)$. 
\end{lemma} 
\begin{proof}
If $(v,y)$ is not an edge in $\LDT(v)$, then the claim follows from 
Lemma~\ref{lem02}. In the rest of the proof, we assume that $(v,y)$ is 
an edge in $\LDT(v)$. We have to show that algorithm $\aPLDG(y)$ removes 
the edge $(v,y)$ from $E(y)$. Thus, we have to show that there exists a 
point $x'$ in $N_y$ which broadcasts the center of the circumcircle of 
some triangular face in $\LDT(x')$ and, based on this information, 
$\aPLDG(y)$ removes $(v,y)$. We will use the edge $(x,q)$ to prove 
that such a point $x'$ exists.  
We assume, without loss of generality, that $vy$ is horizontal and $v$ 
is to the right of $y$. For each $x' \in V \setminus \{v,y\}$, let 
\[ Q_{vy}(x') = \{ q' \in V \setminus \{v,y\} : (x',q') 
                   \mbox{ is an edge in } \LDT(x') 
                   \mbox{ and } 
                   x'q' \mbox{ crosses } vy 
                \} .
\] 
We define 
\[ X_{vy} = \{ x' \in V \setminus \{v,y\} : |x'y| \leq 1 , 
                         |x'v| \leq 1 , Q_{vy}(x') \neq \emptyset 
            \} .
\] 
Since $q \in Q_{vy}(x)$, we have $Q_{vy}(x) \neq \emptyset$. 
Since $|xy| \leq 1$ and $|xv| \leq 1$, we have $x \in X_{vy}$ and, 
therefore, $X_{vy} \neq \emptyset$. 

Let $x'$ be the leftmost point in $X_{vy}$. Let $q'$ be the point in 
$Q_{vy}(x')$ such that the intersection between $x'q'$ and $vy$ is 
closest to $y$. We assume, without loss of generality, that $x'$ is 
above the line through $vy$. Since $x'q'$ crosses $vy$, the point $q'$ 
is below the line through $vy$. Observe that $x'$, $q'$, $v$, and $y$ 
are pairwise distinct. 

\begin{figure}
  \centering
  \includegraphics[scale=0.52]{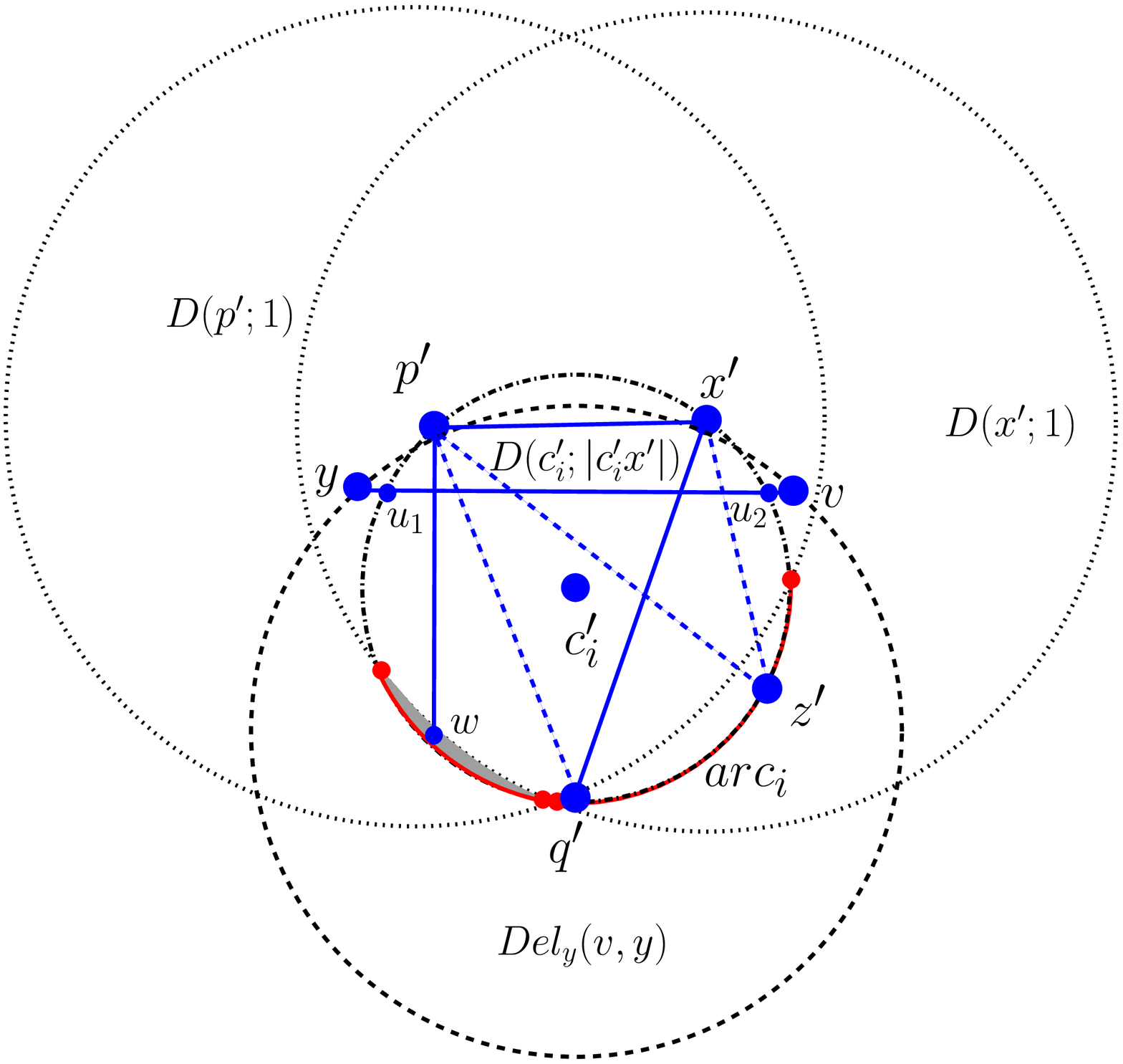}
  \caption{\sl An illustration of the proof of Lemma \ref{lem04}.}  
  \label{fig04}
\end{figure}

By definition, $(x',q')$ is an edge in $\LDT(x')$. Let $p'$ be the point 
of $V$ such that $\Delta(x',p',q')$ is a triangular face in $\LDT(x')$ 
and $p'$ is to the left of the directed line from $q'$ to $x'$; refer to 
Figure \ref{fig04}. Since $y \in N_{x'}$ and $y$ is to the left of this 
line, the point $p'$ exists. Observe that $p'$ may be equal to $y$. 

The following two facts imply that (i) $p'$ is not below the line 
through $vy$, and (ii) in the case when $p' \neq y$, $p'q'$ crosses $vy$: 
First, since $y \in N_{x'}$ and $\Delta(x',p',q')$ is a triangular 
face in $\LDT(x')$, the point $y$ cannot be in $\Delta(x',p',q')$. 
Second, by our choice of $q'$, the line segments $x'p'$ and $vy$ do 
not cross. 
 
In the rest of the proof, we will prove the following two claims: 
\begin{enumerate} 
\item Algorithm $\aPLDG(x')$ broadcasts the center of the circumcircle 
      of $\Delta(x',p',q')$. Since $|x'y| \leq 1$, $y$ will receive 
      this center.    
\item When algorithm $\aPLDG(y)$ considers the center of the 
      circumcircle of $\Delta(x',p',q')$, it deletes the edge $(v,y)$. 
      As a result, the edge $(v,y)$ is not in $E(y)$. 
\end{enumerate}   

Let $c'_i$ be the center of the circumcircle of $\Delta(x',p',q')$ and 
consider the corresponding disk $D(c'_i;|c'_ix'|)$. Recall that 
$D(x';1)$ denotes the unit-disk centered at $x'$. 

Since $|x'y| \leq 1$, $|x'v| \leq 1$, and $\Delta(x',p',q')$ is a 
triangular face in $\LDT(x')$, neither $v$ nor $y$ is contained in the 
interior of $D(c'_i;|c'_ix'|)$. Moreover, since $v \not\in \{x',p',q'\}$, 
$v$ is not contained in $\partial D(c'_i;|c'_ix'|)$. Finally, in the 
case when $y \neq p'$, $y$ is not contained in 
$\partial D(c'_i;|c'_ix'|)$. Since $vy$ crosses $x'q'$, it follows that 
any disk $D$ with $v$ and $y$ on its boundary contains at least one of 
$x'$ and $q'$ (because otherwise, $\partial D$ and 
$\partial D(c'_i;|c'_ix'|)$ intersect more than twice). 

We now show that $|yq'| > 1$. Assume, by contradiction, that 
$|yq'| \leq 1$. Since $(v,y)$ is an edge in $\LDT(y)$, there exists a 
disk $\Del_y(v,y)$ having the property that 
$N_y \cap \partial \Del_y(v,y) = \{v,y\}$ and  
$N_y \cap \Int(\Del_y(v,y)) = \emptyset$. Since both $x'$ and $q'$ are 
in $N_y$, neither of these two points is contained in $\Del_y(v,y)$, 
which is a contradiction. Since $(v,y)$ is an edge in $\LDT(v)$, a 
symmetric argument implies that $|vq'|>1$. 

Consider the triangle $\Delta(v,y,q')$. Since $|vq'| > 1$, $|yq'|>1$, 
and $|vy| \leq 1$, we have $\angle{vq'y} < \frac{\pi}{3}$. Since  
$\angle {x'q'p'} \leq \angle{vq'y}$ it follows that 
$\angle {x'q'p'} < \frac{\pi}{3}$. Next, consider the triangle 
$\Delta(x',p',q')$. Since 
$\angle{x'p'q'} + \angle{p'x'q'} > \frac{2\pi}{3}$, at least one of 
$\angle{x'p'q'}$ and $\angle{p'x'q'}$ is larger than $\frac{\pi}{3}$. 
Below, we will prove that $\angle{p'x'q'} > \frac{\pi}{3}$. 
Since $\Delta(x',p',q')$ is a triangular face in $\LDT(x')$, this will 
imply that algorithm $\aPLDG(x')$ broadcasts a sequence in line~8 which 
contains the center $c'_i$ of $D(c'_i;|c'_ix'|)$.  

Assume, by contradiction, that $\angle{p'x'q'} \leq \frac{\pi}{3}$. 
Then $\angle{x'p'q'} > \frac{\pi}{3}$. Since $|x'p'| \leq 1$, 
$|x'q'| \leq 1$ and $\angle{p'x'q'} \leq \frac{\pi}{3}$, we have 
$|p'q'| \leq 1$. We first prove, again by contradiction, that 
$\Delta(x',p',q')$ is not a triangular face in $\LDT(p')$. Thus, we 
assume that it is a triangular face in $\LDT(p')$. 
Since $vy$ crosses $x'q'$, and $(v,y)$ is an edge in $\LDT(y)$, we 
have $p' \neq y$ (because otherwise, $\LDT(p')$ would not be plane). 
Refering again to Figure \ref{fig04}, let $u_1$ and $u_2$ be the two 
intersection points between $D(c'_i;|c'_ix'|)$ and $vy$, where $u_1$ is 
to the left of $u_2$. We have 
$\angle{u_1 q' u_2} \leq \angle{vq'y} < \frac{\pi}{3}$. 
Consider the arc $\widehat{u_1 x' u_2}$ on $\partial D(c'_i;|c'_ix'|)$ 
with endpoints $u_1$ and $u_2$ that contains $x'$. This arc is a minor 
arc on $\partial D(c'_i;|c'_ix'|)$. Since $p' \in \widehat{u_1 x' u_2}$ 
and $p'$ is to the left of the line through $x'q'$, it follows that 
$p'$ is to the left of $x'$. Then, by our choice of $x'$, we have 
$p' \not\in X_{vy}$. Thus, by the definition of $X_{vy}$, we have 
(i) $|p'y|>1$ or (ii) $|p'v|>1$ or (iii) $Q_{vy}(p') = \emptyset$. 
Since $q' \in Q_{vy}(p')$, (iii) does not hold. Since $vy$ and $p'q'$ 
cross, $|vy| \leq 1$, $|p'q'| \leq 1$, $|yq'|>1$, and $|vq'|>1$, it 
follows from Lemma~\ref{lem01} that $|p'y| \leq 1$ and $|p'v| \leq 1$; 
thus, neither (i) nor (ii) holds, which is a contradiction. We conclude 
that $\Delta(x',p',q')$ is not a triangular face in $\LDT(p')$. 

We continue deriving a contradiction to the assumption that 
$\angle{p'x'q'} \leq \frac{\pi}{3}$. Since $\Delta(x',p',q')$ is a 
triangular face in $\LDT(x')$ but not in $\LDT(p')$, there exists at 
least one point $w$ of $V$ in the interior of $D(c'_i;|c'_ix'|)$ such 
that $|x'w|>1$ and $|p'w| \leq 1$. Let $W$ be the set of all such points 
$w$, i.e., 
\[ W = \{ w \in V : w \in \Int(D(c'_i;|c'_ix|)), |x'w|>1 , |p'w| \leq 1 
       \} . 
\] 
For each $w \in W$, let $R_w$ be the radius of the circle through $p'$ 
and $w$ and whose center is on $p'c'_i$. Let $w$ be a point in $W$ for 
which $R_w$ is minimum. Let $D_w$ be the disk centered on $p'c'_i$ that 
contains $p'$ and $w$ on its boundary. Observe that $D_w$ is contained 
in $D(c'_i;|c'_ix'|)$. Also, no point of $W$ is in the interior of 
$D_w$. It follows that $(p',w)$ is an edge in $\LDT(p')$. 

We have seen above that $p'$ is to the left of $x'$. It follows that 
$p' \not\in X_{vy}$. Thus, by the definition of $X_{vy}$, we have 
(i) $|p'y|>1$, or (ii) $|p'v|>1$, or (iii) $Q_{vy}(p') = \emptyset$, or 
(iv) $p' = y$. The arguments above show that neither (i) nor (ii) holds. 
Assume that (iv) does not hold, i.e., $p' \neq y$. We show that 
$w \in Q_{vy}(p')$; this will imply that (iii) does not hold. 
Since $w \in \Int(D(c'_i;|c'_ix|))$, we have $w \neq v$ and 
$w \neq y$. Thus, in order to show that $w \in Q_{vy}(p')$, it suffices 
to show that $p'w$ crosses $vy$. Consider again the two intersection 
points $u_1$ and $u_2$ between $D(c'_i;|c'_ix'|)$ and $vy$, where $u_1$ 
is to the left of $u_2$. As we have seen before, the arc 
$\widehat{u_1 x' u_2}$ on $\partial D(c'_i;|c'_ix'|)$ is a minor arc. 
Since $|u_1 u_2| \leq |vy| \leq 1$, it follows from Lemma~\ref{lem03} 
that $w$ is below the line through $v$ and $y$ (because otherwise, we 
would have $|wx'| \leq 1$, contradicting the fact that $w \in W$). Thus, 
since $p'$ and $w$ are on opposite sides of the line through $v$ and 
$y$, and since $w \in D(c'_i;|c'_ix'|)$, this shows that $p'w$ crosses 
$vy$. As mentioned above, this implies that (iii) does not hold. We 
conclude that (iv) holds, i.e., $p'=y$. In the triangle 
$\Delta(p',x',q')$, we have $|p'x'| \leq 1$, $|x'q'| \leq 1$, and 
$|p'q'|=|yq'|>1$. It follows that $\angle{p'x'q'} > \frac{\pi}{3}$, 
which is a contradiction. 

Thus, we have obtained a contradiction to the assumption that 
$\angle{p'x'q'} \leq \frac{\pi}{3}$. As a result, we conclude that   
$\angle{p'x'q'} > \frac{\pi}{3}$. As we mentioned before, this implies 
that algorithm $\aPLDG(x')$ broadcasts a sequence in line~8 which 
contains the center $c'_i$ of $D(c'_i;|c'_ix'|)$ (which is the 
circumscribing disk of the triangular face $\Delta(x',p',q')$ in 
$\LDT(x')$). 

Since $|yx'| \leq 1$, $y$ receives the sequence broadcast by algorithm  
$\aPLDG(x')$. This sequence contains the center $c'_i$ together with 
the point $x'$. Recall that $|yq'|>1$. Let $D$ be the disk whose 
boundary contains $y$, $v$, and the ``north pole'' of $D(c'_i;|c'_ix'|)$. 
Since $\angle{vq'y} < \frac{\pi}{3}$, the center of $D$ is below the 
line through $v$ and $y$. Since $|vy| \leq 1$, it then follows from 
Lemma~\ref{lem03} that $|yp'| \leq 1$. Thus, when algorithm $\aPLDG(y)$ 
considers $c'_i$, it discovers that the boundary of the disk 
$D(c'_i;|c'_ix'|)$ contains exactly two points of $N_y$; these are the 
points $x'$ and $p'$. Algorithm $\aPLDG(y)$ computes the open arc 
$\arc_i$, which is the part of $\partial D(c'_i;|c'_ix'|)$ that is not 
contained in the unit-disk $D(y;1)$ centered at $y$. The algorithm knows 
that the third point $q'$ on $\partial D(c'_i;|c'_ix'|)$ is somewhere 
on $\arc_i$. Let $\Del_y(v,y)$ be the disk that is computed in line~17 
of algorithm $\aPLDG(y)$. This disk has the properties that 
$N_y \cap \partial \Del_y(v,y) = \{v,y\}$ and 
$N_y \cap \Int(\Del_y(v,y)) = \emptyset$. 
By the same argument as in the proof of Lemma~\ref{lem02}, $\arc_i$ is 
contained in the interior of $Del_y(v,y)$. Moreover, $\arc_i$ is 
below the line through $v$ and $y$. 

Consider the point $z'$ on $\arc_i$ that is chosen in line~15 of 
algorithm $\aPLDG(y)$. We will show that $vy$ crosses $x'z'$. 
Assume, by contradiction, that $vy$ does not cross $x'z'$. Since the 
line through $v$ and $y$ separates $x'$ from the two points $q'$ and 
$z'$, and since $vy$ crosses $x'q'$, it follows that $v$ or $y$ is in 
the triangle $\Delta(x',q',z')$. Thus, $v$ or $y$ is in the interior 
of the disk $D(c'_i;|c'_i x'|)$. This is a contradiction, because  
$|x'v| \leq 1$, $|x'y| \leq 1$, and $\Delta(x',p',q')$ is a triangular 
face in $\LDT(x')$. Thus, we have shown that $vy$ crosses $x'z'$. 

It now follows from the description of the algorithm that $\aPLDG(y)$ 
removes the edge $(v,y)$ from $E(y)$. This completes the proof of the 
lemma. 
\end{proof}

Based on the previous lemmas, we can now prove that $\PLDG(V)$ is plane: 

\begin{lemma}\label{lem05}
       $\PLDG(V)$ is a plane graph. 
\end{lemma}
\begin{proof}
The proof is by contradiction. Assume that $\PLDG(V)$ contains two 
crossing edges $(v,y)$ and $(x,q)$. By Lemma~\ref{lem01}, one of the 
points in $\{x,q,v,y\}$ is within distance $1$ from the other three 
points. We may assume without loss of generality that $|xq| \leq 1$, 
$|xv| \leq 1$, and $|xy| \leq 1$. By Lemma~\ref{lem02}, $(v,y)$ is an 
edge in $\LDT(v)$ and in $\LDT(y)$, and $(x,q)$ is an edge in $\LDT(x)$. 

Since all conditions in Lemma~\ref{lem04} are satisfied, $(v,y)$ is not 
an edge in $E(y)$. Also, the conditions in Lemma~\ref{lem04}, with 
$v$ and $y$ interchanged, are satisfied. Therefore, $(v,y)$ is not an 
edge in $E(v)$. Thus, $(v,y)$ is not an edge in $\PLDG(V)$, which is 
a contradiction.  
\end{proof} 

The following lemma summarizes the different scenarios when algorithm 
$\aPLDG(v)$ removes an edge $(v,y)$ from the edge set $E(v)$.

\begin{lemma}\label{lem06} 
       Let $v$ and $y$ be two distinct points of $V$ such that $(v,y)$ 
       is an edge in $\LDT(v)$. Assume that algorithm $\aPLDG(v)$ 
       removes $(v,y)$ from $E(v)$. Then, there exist three pairwise 
       distinct points $x$, $p$, and $q$ in $V$ such that 
       \begin{enumerate} 
       \item $\Delta(x,p,q)$ is a triangular face in $\LDT(x)$, 
       \item $v \neq x$, $|vx| \leq 1$, $|vp| \leq 1$, $|vq| > 1$, 
       \item neither $v$ nor $y$ is in the interior of the disk 
             $D(x,p,q)$, and  
       \item \begin{enumerate} 
             \item if $y \neq x$, $v \neq p$, and $y \neq p$, the line 
                   segment $vy$ crosses both the line segments $xq$ 
                   and $pq$, 
             \item if $y=x$, the line segment $vy$ crosses the line 
                   segment $pq$, 
             \item if $v=p$, the line segment $vy$ crosses the line 
                   segment $xq$, 
             \item if $y=p$, the line segment $vy$ crosses the line 
                   segment $xq$. 
             \end{enumerate} 
       \end{enumerate} 
 \end{lemma}
\begin{proof} 
Since algorithm $\aPLDG(v)$ removes $(v,y)$ from $E(v)$, there exists a 
point $x$ in $N_v \setminus \{v\}$ which broadcasts the center $c'_i$ 
of the circumcircle of a triangular face $\Delta(x,p,q)$ in $\LDT(x)$, 
such that the following holds:
\begin{enumerate} 
\item Consider the disk $D(c'_i;|c'_ix|) = D(x,p,q)$ with center $c'_i$ 
      that contains $x$, $p$, and $q$ on its boundary. Then, according 
      to line~12 of algorithm $\aPLDG(v)$, $\partial D(c'_i;|c'_ix|)$ 
      contains exactly two points of $N_v$. Since we assume that no 
      four points of $V$ are cocircular, $x$, $p$, and $q$ are the only 
      points of $V$ that are on $\partial D(c'_i;|c'_ix|)$. 
      Thus, since $x \in N_v$, exactly one of $p$ and $q$ is in $N_v$. 
      We may assume without loss of generality that $p \in N_v$ and 
      $q \not\in N_v$. 
\item Consider the unit-disk $D(v;1)$ centered at $v$. Let $\arc_i$ be 
      the arc on $\partial D(c'_i;|c'_ix|)$ that is not contained in 
      $D(v;1)$, let $z'$ be the point on $\arc_i$ with $|xz'| \leq 1$ 
      or $|pz'| \leq 1$ that is chosen by algorithm $\aPLDG(v)$ in 
      line~15, and let $\Del_v(v,y)$ be the disk chosen in line~17. 
      Thus, $v$ and $y$ are the only points of $N_v$ that are on 
      $\partial \Del_v(v,y)$ and no point of $N_v$ is in the interior 
      of $\Del_v(v,y)$. Then, by line~18 of algorithm $\aPLDG(v)$, 
      $\arc_i$ is contained in the interior of $\Del_v(v,y)$ and $vy$ 
      crosses at least one of $xz'$ and $pz'$. 
\end{enumerate}  
The first two claims in the lemma hold for the points $x$, $p$, and $q$. 

We now prove the third claim. Since $|xv| \leq 1$ and $\Delta(x,p,q)$ 
is a triangular face in $\LDT(x)$, $v$ is not in the interior of the disk 
$D(x,p,q)$. We prove by contradiction that $y$ is not in the interior 
of $D(x,p,q)$. Thus, we assume that $y$ is in the interior of this disk. 
Then, again since $\Delta(x,p,q)$ is a triangular face in $\LDT(x)$, 
we have $|xy|>1$. Recall that (i) $|xz'| \leq 1$ or $|pz'| \leq 1$ and 
(ii) $vy$ crosses at least one of $xz'$ and $pz'$. Therefore, we 
distinguish four cases and derive a contradiction for each of them. 

\vspace{0.5em} 

\noindent 
{\bf Case 1:} $|xz'| \leq 1$ and $vy$ crosses $xz'$. 

Since $|vy| \leq 1$ and $|vz'|>1$, Lemma~\ref{lem01} implies that 
$|xy| \leq 1$, which is a contradiction. 

\vspace{0.5em} 

\noindent 
{\bf Case 2:} $|xz'| \leq 1$ and $vy$ does not cross $xz'$. 

In this case, $vy$ crosses $pz'$. Observe that $x \neq y$, $v \neq p$, 
and $y \neq p$. Also, $v \not\in D(x,p,q)$. The following observations 
lead to a contradiction: 
\begin{itemize} 
\item Since $|xp| \leq 1$ and $|xz'| \leq 1$, each point in the triangle 
      $\Delta(x,p,z')$ has distance at most one to $x$. Therefore, 
      $y \not\in \Delta(x,p,z')$. 
\item The line segment $vy$ crosses $px$. This follows from the facts 
      that $vy$ does not cross $xz'$, $vy$ crosses $pz'$, 
      $v \not\in D(x,p,q)$, $y \in \Int(D(x,p,q))$, and 
      $y \not\in \Delta(x,p,z')$. 
\item The line segment $px$ is disjoint from $\arc_i$: Since 
      $|vp| \leq 1$ and $|vx| \leq 1$, each point on $px$ has distance 
      at most one to $v$. Thus, $px \subseteq D(v;1)$. However, 
      $\arc_i$ and $D(v;1)$ are disjoint. 
\item $\arc_i$ and $v$ are on the same side of the line through $p$ and 
      $x$: Assume this is not the case. Since neither $p$ nor $x$ is 
      in $\Del_v(v,y)$ and since $\arc_i$ is in the interior of 
      $\Del_v(v,y)$, it follows that $\partial \Del_v(v,y)$ and 
      $\partial D(x,p,q)$ intersect more than twice. This is a 
      contradiction. 
\item Let $\widehat{px}$ be the arc on $\partial D(x,p,q)$ between $p$ 
      and $x$ that does not contain $z'$. We claim that $\widehat{px}$ 
      is a major arc. To prove this, assume that it is a minor arc. 
      The observations above imply that $y$ is in the region of 
      $D(x,p,q)$ that is bounded by $px$ and $\widehat{px}$. Since 
      $|xp| \leq 1$, it then follows from Lemma~\ref{lem03} that 
      $|xy| \leq 1$, which is a contradiction. 
\item Let $v'$ be the intersection between $xv$ and $\partial D(x,p,q)$. 
      Let $\widehat{pv'x}$ be the arc on $\partial D(x,p,q)$ between 
      $p$ and $x$ that contains $v'$. Since $|vp| \leq 1$, 
      $|vx| \leq 1$, and $\widehat{pv'x}$ is a minor arc, we know that 
      $\widehat{pv'x}$ is contained in $D(v;1)$. Since $q$ is on 
      $\widehat{pv'x}$, it follows that $|vq| \leq 1$, which is a 
      contradiction. 
\end{itemize} 

\vspace{0.5em} 

\noindent
{\bf Case 3:} $|xz'|>1$ and $vy$ crosses $xz'$. 

The following observations lead to a contradiction: 
\begin{itemize} 
\item Since $|vy| \leq 1$, $|xq| \leq 1$, $|xy|>1$, and $|vq|>1$, it 
      follows from Lemma~\ref{lem01} that $vy$ and $xq$ do not cross. 
      This also implies that $q \neq z'$. 
\item The points $q$ and $z'$ are on the same side of the line through 
      $v$ and $y$: This follows from the facts that both $q$ and $z'$ 
      are on $\arc_i$, $\arc_i$ is contained in the interior of 
      $\Del_v(v,y)$, $\arc_i \cap D(v;1) = \emptyset$, and 
      $|vy| \leq 1$. 
\item Since $vy$ crosses $xz'$ but not $xq$, and since $q$ and $z'$ 
      are on the same side of the line through $v$ and $y$, it follows 
      that $y$ is in the triangle $\Delta(x,q,z')$. 
\item Consider the unit-disk $D(x;1)$ centered at $x$. Assume, without 
      loss of generality that $xq$ is vertical, $q$ is above $x$, and 
      $z'$ is to the right of $x$. Observe that both $v$ and $q$ are 
      contained in $D(x;1)$, and neither $y$ nor $z'$ is contained in 
      $D(x;1)$. Since $vy$ crosses $xz'$ and $y \in \Delta(x,q,z')$, 
      the point $v$ is below the line through $x$ and $z$. This implies 
      that the line through $v$ and $y$ separates $q$ and $z'$, which 
      is a contradiction. 
\end{itemize} 

\vspace{0.5em} 

\noindent 
{\bf Case 4:} $|xz'|>1$ and $vy$ does not cross $xz'$. 

In this case, $vy$ crosses $pz'$. The following observations lead to 
a contradiction: 

\begin{itemize} 
\item The line segments $vy$ and $px$ do not cross: If they do cross, 
      then the same analysis as in Case~2 leads to a contradiction. 
\item As in Case~3, the line segments $vy$ and $qx$ do not cross. 
\item Since $vy$ crosses $pz'$, but $vy$ neither crosses $xz'$ nor $xp$, 
      and since $y \in \Int(D(x,p,q)$ and $v \not\in \Int(D(x,p,q))$, 
      the point $y$ is in the triangle $\Delta(x,p,z')$. 
\item Since $|xp| \leq 1$ and $|xq| \leq 1$, each point in the triangle 
      $\Delta(x,p,q)$ has distance at most one to $x$. Therefore, 
      since $|xy|>1$, we have $y \not\in \Delta(x,p,q)$. In particular, 
      $q \neq z'$.  
\item As in Case~2, the line segment $px$ is disjoint from $\arc_i$.  
      Thus, $q$ and $z'$ are on the same side of the line through $p$ 
      and $x$. 
\item Assume without loss of generality that $px$ is horizontal, $p$ is 
      to the left of $x$, and both $q$ and $z'$ are above the line 
      through $p$ and $x$. 
\item Let $\widehat{pqx}$ be the arc on $\partial D(x,p,q)$ that is above 
      $px$. If this arc is a minor arc, then, since $|px| \leq 1$ and 
      using Lemma~\ref{lem03}, we have $|xz'| \leq 1$, which is a 
      contradiction. Thus, $\widehat{pqx}$ is a major arc. 
\item Assume that $y$ is on or below $px$. Since the arc on 
      $\partial D(x,p,q)$ that is below $px$ is a minor arc, it follows 
      from Lemma~\ref{lem03} that $|xy| \leq 1$, which is a 
      contradiction. Thus, $y$ is above $px$. 
\item Assume that $y$ is to the right of $xq$. Since $y$ is contained 
      in $\Delta(x,p,z')$, the point $z'$ is on the arc on 
      $\partial D(x,p,q)$ between $x$ and $q$ that is to the right of 
      $xq$. Recall that $vy$ crosses neither $xz'$ nor $xq$. It follows 
      that $vy$ crosses $qz'$, which is a contradiction, because 
      $q$ and $z'$ are on the same side of the line through $v$ and 
      $y$. 
\item We conclude that $y$ is to the left of $xq$. Since $y$ is 
      contained in $\Delta(x,p,z')$ but not in $\Delta(x,p,q)$, the 
      point $z'$ is on the arc on $\partial D(x,p,q)$ between $p$ and 
      $q$ that is to the left of $xq$. 
\item Assume that $v$ is above the line through $x$ and $y$. Since $vy$ 
      crosses $pz'$, $y$ is contained in the triangle $\Delta(x,p,v)$. 
      However, since $|xv| \leq 1$ and $|xp| \leq 1$, this implies 
      that $|xy| \leq 1$, which is a contradiction. Thus, $v$ is below 
      the line through $x$ and $y$. 
\item Since both $q$ and $z'$ are above the line through $v$ and $y$, 
      $y$ is contained in the triangle $\Delta(x,v,q)$. However, since 
      $|xv| \leq 1$ and $|xq| \leq 1$, this implies that $|xy| \leq 1$, 
      which is a contradiction. 
\end{itemize} 

To conclude, in each of the four cases above, we have obtained a 
contradiction to the assumption that $y$ is in the interior of 
$D(x,p,q)$. Therefore, we have proved the third claim in the lemma. 

It remains to prove the fourth claim in the lemma. First assume that 
$y \neq x$, $v \neq p$, and $y \neq p$. We first show that $x$ and 
$p$ are on the same side of the line through $v$ and $y$. Assume, by 
contradiction, that $x$ and $p$ are on opposite sides of this line. 
Since both $x$ and $p$ are in $N_v$, neither of these two points is 
contained in $\Del_v(v,y)$. On the other hand, since 
$\arc_i \subseteq \Int(\Del_v(v,y))$ and $z' \in \arc_i$, the 
point $z'$ is in the interior of $\Del_v(v,y)$. We also know that 
neither $v$ nor $y$ is contained in $D(c'_i;|c'_i x|) = D(x,p,q)$. 
Since $\partial D(x,p,q)$ contains the points $x$, $p$, and $z'$,  
it follows that the boundaries of $\Del_v(v,y)$ and $D(x,p,q)$ intersect 
more than twice. This is a contradiction. 
 
Assume, without loss of generality, that $vy$ is horizontal and both 
$x$ and $p$ are above the line through $v$ and $y$. Then $z'$ is 
below this line. Since $\arc_i \cap D(v;1) = \emptyset$, it follows 
that the entire arc $\arc_i$ is below this line. In particular, $q$ is 
below the line through $v$ and $y$. Since neither $v$ nor $y$ is 
contained in $D(x,p,q)$, since $vy$ intersects $\partial D(x,p,q)$ 
twice, and since $vy$ separates $q$ from $x$ and $p$, it follows that 
$vy$ crosses both the line segments $xq$ and $pq$. 

It remains to prove the special cases in the fourth claim. 
First assume that $y=x$. Since $vy$ does not cross $xz' = yz'$, we know 
that $vy$ crosses $pz'$, which implies that $v \neq p$ and $y \neq p$. 
Since the line through $v$ and $y$ separates $p$ from $q$ and $z'$, 
it follows that $vy$ crosses $pq$. 

Next assume that $v=p$. Since $vy$ does not cross $pz' = vz'$, we know 
that $vy$ crosses $xz'$, which implies that $y \neq x$. Since $q$ and 
$z'$ are on the same side of the line through $v$ and $y$, it 
follows that $vy$ crosses $xq$. 

Finally, assume that $y=p$. Since $vy$ does not cross $pz' = yz'$, we 
know that $vy$ crosses $xz'$. Since the line through $v$ and $y$ 
separates $x$ from $q$ and $z'$, it follows that $vy$ crosses $xq$. 
This completes the proof of the lemma.  
\end{proof} 

We can now prove that $\PLDG(V)$ is consistent: 
 
\begin{lemma}     \label{lem07}
       The graph $\PLDG(V)$ is consistent: For any two distinct points 
       $v$ and $y$ of $V$, $(v,y)$ is an edge in $E(v)$ if and only if 
       $(v,y)$ is an edge in $E(y)$. 
\end{lemma}
\begin{proof}
The proof is by contradiction. Assume there is a pair $(v,y)$ which is 
an edge in $E(y)$ but not in $E(v)$. Then $(v,y)$ is an edge in $\LDT(y)$ 
and, by Lemma~\ref{lem02}, $(v,y)$ is an edge in $\LDT(v)$. Since 
$(v,y)$ is not an edge in $E(v)$, it has been removed by algorithm 
$\aPLDG(v)$. Thus, by Lemma~\ref{lem06}, there exist three pairwise 
distinct points $x$, $p$, and $q$ in $V$ such that  
(i) $\Delta(x,p,q)$ is a triangular face in $\LDT(x)$, 
(ii) $v \neq x$, $|vx| \leq 1$, $|vq| > 1$, and 
(iii) the line segment $vy$ crosses at least one of the line segments 
$pq$ and $xq$. 

Assume that $vy$ does not cross $xq$. Then $vy$ crosses $pq$ and, by
the fourth claim in Lemma~\ref{lem06}, $y=x$. Thus, since 
$(v,y)$ is an edge in $LDT(y) = \LDT(x)$ and using (i), it follows 
that $\LDT(x)$ is not plane, which is a contradiction. 

Thus, $vy$ crosses $xq$. This implies that the points $x$, $q$, $v$, 
and $y$ are pairwise distinct. It follows from (i) that $(x,q)$ is an 
edge in $\LDT(x)$ and $|xq| \leq 1$. Since $|vy| \leq 1$, $|xq| \leq 1$, 
$|vq|>1$, and since $vy$ crosses $xq$, it follows from Lemma~\ref{lem01} 
that $|xy| \leq 1$. Thus, all conditions in Lemma~\ref{lem04} are 
satisfied. As a result, algorithm $\aPLDG(y)$ deletes the edge $(v,y)$ 
from $E(y)$. This is a contradiction. 
\end{proof} 

Recall that $\UDel(V)$ denotes the intersection of the Delaunay 
triangulation and the unit-disk graph of $V$. We next show that 
$\PLDG(V)$ contains $\UDel(V)$.  

\begin{lemma} \label{lem08}
       The graph $\UDel(V)$ is a subgraph of $\PLDG(V)$. 
\end{lemma}
\begin{proof} 
Let $(v,y)$ be an edge of $\UDel(V)$. We will show that $(v,y)$ is an 
edge in $E(v)$. By definition, $|vy| \leq 1$ and $(v,y)$ is an edge in 
the Delaunay triangulation of $V$. Therefore, $(v,y)$ is also an edge in 
the Delaunay triangulation $\LDT(v)$ of $N_v$ and, thus, $(v,y)$ is 
added to the edge set $E(v)$ in line~3 of algorithm $\aPLDG(v)$. 
We have to show that algorithm $\aPLDG(v)$ does not remove $(v,y)$ in 
line~19. 

Assume that $(v,y)$ is removed in line~19 of algorithm $\aPLDG(v)$.
By Lemma~\ref{lem06}, there exist three pairwise distinct points $x$, 
$p$, and $q$ in $V$ such that (i) neither $v$ nor $y$ is in the interior 
of the disk $D(x,p,q)$ and (ii) the line segment $vy$ crosses at least 
one of the line segments $xq$ and $pq$.

Assume that $vy$ crosses $xq$. Then, the points $v$, $y$, $x$, and $q$ 
are pairwise distinct. Observe that $p$ may be equal to $v$ or $y$. 
Let $D$ be an arbitrary disk having $v$ and $y$ on its boundary, and 
assume that neither $x$ nor $q$ is contained in $D$. Then it follows 
from (i) and (ii) that the boundaries of $D$ and $D(x,p,q)$ intersect 
more than twice, which is a contradiction. Thus, $D$ contains at least 
one of $x$ and $q$. Since $D$ was arbitrary, this contradicts the fact 
that $(v,y)$ is an edge in the Delaunay triangulation of $V$.  

By a symmetric argument, the case when $vy$ crosses $pq$ also leads to 
a contradiction to the fact that $(v,y)$ is an edge in the Delaunay 
triangulation of $V$.  
\end{proof} 

In the next lemma, we summarize the results obtained in this section. 
Recall that a message is defined to be the location of a point in the 
plane. 

\begin{lemma}   \label{lem10}
       Let $V$ be a finite set of points in the plane. The distributed 
       algorithm $\aPLDG(v)$, where $v$ ranges over all points in $V$, 
       is a $2$-local algorithm that computes a plane and consistent  
       $\frac{4 \pi \sqrt{3}}{9}$-spanner $\PLDG(V)$ of the unit-disk 
       graph of $V$. This algorithm makes one communication round and 
       each point of $V$ broadcasts at most $6$ messages.  
\end{lemma}
\begin{proof} 
Let $v$ be a point of $V$. Lines 1--8 of algorithm $\aPLDG(v)$ depend 
only on the points in $N_v$. Lines 9--19 depend only on information 
received from nodes $x$ in $N_v$; this information was computed in 
lines 1--8 of algorithm $\aPLDG(x)$ and, thus, depends only on the 
points in $N_x$. It follows that algorithm $\aPLDG(v)$ is 2-local.  
Algorithm $\aPLDG(v)$ broadcasts a sequence of messages only once, 
in line~8. Therefore, there is only one round of communication. 
In the Delaunay triangulation $\LDT(v)$ of $N_v$, there are at most 
$5$ triangular faces $\Delta(u,v,w)$ with $\angle{uvw} >\frac{\pi}{3}$. 
Therefore, the sequence that is broadcast in line~8 contains at most 
$6$ points. Thus, $\aPLDG(v)$ broadcasts at most $6$ messages.  

By Lemmas~\ref{lem05} and~\ref{lem07}, the graph $\PLDG(V)$ is plane 
and consistent. By Lemma~\ref{lem08}, $\PLDG(V)$ is a supergraph of 
$\UDel(V)$. Since $\UDel(V)$ is a $\frac{4 \pi \sqrt{3}}{9}$-spanner 
of the unit-disk graph $\UDG(V)$ of $V$ (see 
Bose \emph{et al.}~\cite{bmnsz-agbsp-03}), the graph $\PLDG(V)$ is a 
$\frac{4 \pi \sqrt{3}}{9}$-spanner of $\UDG(V)$. 
\end{proof}

\section{The Final Algorithm} \label{improve} 
We have seen that in algorithm $\aPLDG$, each point of $V$ broadcasts 
at most $6$ messages. In this section, we improve this upper bound to 
$5$. We obtain this improvement, by making the following modification to 
the algorithm: The sequence $(v,c_1,\ldots,c_k)$ that is broadcast in 
line~8 of algorithm $\aPLDG(v)$ contains the location of the sender $v$. 
In our new algorithm, point $v$ sends only the sequence 
$(c_1,\ldots,c_k)$ of centers. Thus, any point that receives this 
sequence does not know that the sequence was broadcast by $v$. Assume 
that $v$ receives a center $c'_i$ from some node $x$ in $N_v$. Since 
$v$ does not know that $c'_i$ was broadcast by $x$, line~11 in 
algorithm $\aPLDG(v)$ has to be modified. In the new algorithm, $v$ 
computes a point $x'$ in $N_v \setminus \{v\}$ that is closest to 
$c'_i$ and uses the disk $D(c'_i;|c'_ix'|)$ to decide whether or not 
to remove an edge $(v,y)$.  

The new algorithm, which we denote by $\aPLDG'$, is given in 
Figure~\ref{figPLDG'}. We denote by $E'(v)$ the edge set that is 
computed by algorithm $\aPLDG'(v)$. Let $E' = \cup_{v \in V} E'(v)$ and 
let $\PLDG'(V)$ denote the graph with vertex set $V$ and edge set $E'$. 

\begin{figure}
\begin{algorithm}{{\sc PLDG}$'$}[v]{\label{algo1'}}
let $N_v = \{ u \in V : |uv| \leq 1 \}$;\\
compute the Delaunay triangulation $\LDT(v)$ of $N_v$;\\
let $E'(v)$ be the set of all edges in $\LDT(v)$ that are incident 
on $v$;\\
let $\Delta_v$ be the set of all triangular faces $\Delta(u,v,w)$ in 
$\LDT(v)$ for which $\angle{uvw} > \frac{\pi}{3}$;\\ 
let $k$ be the number of elements in $\Delta_v$;\\ 
\qif $k \geq 1$\\
\qthen let $c_1,\ldots,c_k$ be the centers of the circumcircles of all 
       triangles in $\Delta_v$;\\
       broadcast the sequence $(c_1, \ldots, c_k)$;
\qendif\\
\qfor each sequence $(c'_1, \ldots, c'_m)$ received\\
\qdo \qfor $i=1$ \qto $m$\\
     \qdo compute a point $x'$ in $N_v \setminus \{v\}$ that is closest 
          to $c'_i$;\\
          let $D(c'_i;|c'_i x'|)$ be the disk with center $c'_i$ that 
          contains $x'$ on its boundary;\\
          \qif $\partial D(c'_i;|c'_i x'|)$ contains exactly two points 
               of $N_v$ \\
          \qthen let $p'$ be the point in 
                 $(N_v \setminus \{x'\}) \cap 
                        \partial D(c'_i;|c'_i x'|)$;\\  
                 let $\arc_i = ( \partial D(c'_i;|c'_i x'|) ) 
                                  \setminus D(v;1)$;\\ 
                 let $Z = \{ z' \in \arc_i : |x'z'| \leq 1 \mbox{ or } 
                               |p'z'| \leq 1 \}$;\\ 
                 \qif $\arc_i \neq \emptyset$ and $Z \neq \emptyset$ \\
                 \qthen let $z'$ be an arbitrary element of $Z$;\\ 
                        \qfor each edge $(v,y)$ in $E'(v)$\\ 
                        \qdo let $\Del_v(v,y)$ be a disk $D$ such that 
                             $N_v \cap \partial D = \{v,y\}$ and 
                             $N_v \cap \Int(D) = \emptyset$;\\ 
                             \qif $\arc_i$ is contained in the interior 
                                  of $\Del_v(v,y)$ and the line segment 
                                  $vy$ crosses at least one of the line 
                                  segments $x'z'$ and $p'z'$\\
                             \qthen remove $(v,y)$ from $E'(v)$ 
                             \qendif
                        \qendfor
                 \qendif
          \qendif
     \qendfor
\qendfor 
\end{algorithm}
\caption{\sl The improved plane localized Delaunay graph algorithm.}
\label{figPLDG'}
\end{figure}

Recall that $E(v)$ denotes the edge set that is computed by algorithm 
$\aPLDG(v)$ and $\PLDG(V)$ denotes the graph with vertex set $V$ and 
edge set $\cup_{v \in V} E(v)$. We claim that $\PLDG(V) = \PLDG'(V)$; 
thus, the new algorithm $\aPLDG'$ computes the same graph as algorithm 
$\aPLDG$. In order to prove this claim, it suffices to show that 
algorithm $\aPLDG(v)$ removes an edge $(v,y)$ from $E(v)$ if and only if 
algorithm $\aPLDG'(v)$ removes the edge $(v,y)$ from $E'(v)$. 
We will show this in the following two lemmas. 

\begin{lemma}    \label{lem09}  
       Let $v$ be an element of $V$ and let $(v,y)$ be an edge of the 
       Delaunay triangulation $\LDT(v)$ of the set $N_v$. If algorithm 
       $\aPLDG(v)$ removes $(v,y)$ from $E(v)$, then algorithm 
       $\aPLDG'(v)$ removes $(v,y)$ from $E'(v)$. 
\end{lemma}
\begin{proof} 
By Lemma~\ref{lem06}, there exist three pairwise distinct points 
$x$, $p$, and $q$ in $V$ such that 
\begin{enumerate}  
\item $\Delta(x,p,q)$ is a triangular face in $\LDT(x)$, 
\item $v \neq x$, $|vx| \leq 1$, $|vp| \leq 1$, $|vq| > 1$, 
\item neither $v$ nor $y$ is in the interior of the disk $D(x,p,q)$. 
\end{enumerate}
In fact, in algorithm $\aPLDG(v)$, $v$ receives from $x$ the center 
$c'_i$ of the disk $D(c'_i;|c'_i x|) = D(x,p,q)$. 
Since $|vx| \leq 1$, in algorithm $\aPLDG'(v)$, $v$ receives the center 
$c'_i$, but does not know that it was broadcast by $x$. Consider the 
point $x'$ that is computed in line~11 of algorithm $\aPLDG'(v)$. 
Thus, $x'$ is a point in $N_v \setminus \{v\}$ that is closest to 
$c'_i$. Since $x \in N_v \setminus \{v\}$, we have 
$|c'_i x'| \leq |c'_i x|$. We claim that $|c'_i x'| = |c'_i x|$.  

To prove this claim, assume, by contradiction, that 
$|c'_i x'| < |c'_i x|$. Then $x'$ is in the interior of the disk 
$D(x,p,q)$. Since $\Delta(x,p,q)$ is a triangular face in $\LDT(x)$, 
we have $|xx'| > 1$. 

Consider the disk $\Del_v(v,y)$ that is computed in line~17 of 
algorithm $\aPLDG(v)$. Recall that 
$N_v \cap \partial \Del_v(v,y) = \{v,y\}$ and 
$N_v \cap \Int(\Del_v(v,y)) = \emptyset$. 
Since both $x$ and $x'$ are in $N_v$, neither of these two points is 
in the interior of $\Del_v(v,y)$. It follows from line~18 of 
algorithm $\aPLDG(v)$ that $q$ is in the interior of $\Del_v(v,y)$ 
(because $q \in \arc_i$). 

Assume, without loss of generality that $vy$ is horizontal, $v$ is 
to the right of $y$, and $q$ is below the line through $v$ and $y$; 
refer to Figure~\ref{fig08}. 

\begin{figure}
  \subfigure[$y \neq x$]{
    \label{fig08:a}
    \begin{minipage}[b]{0.5\textwidth}
      \centering
      \includegraphics[scale=0.50]{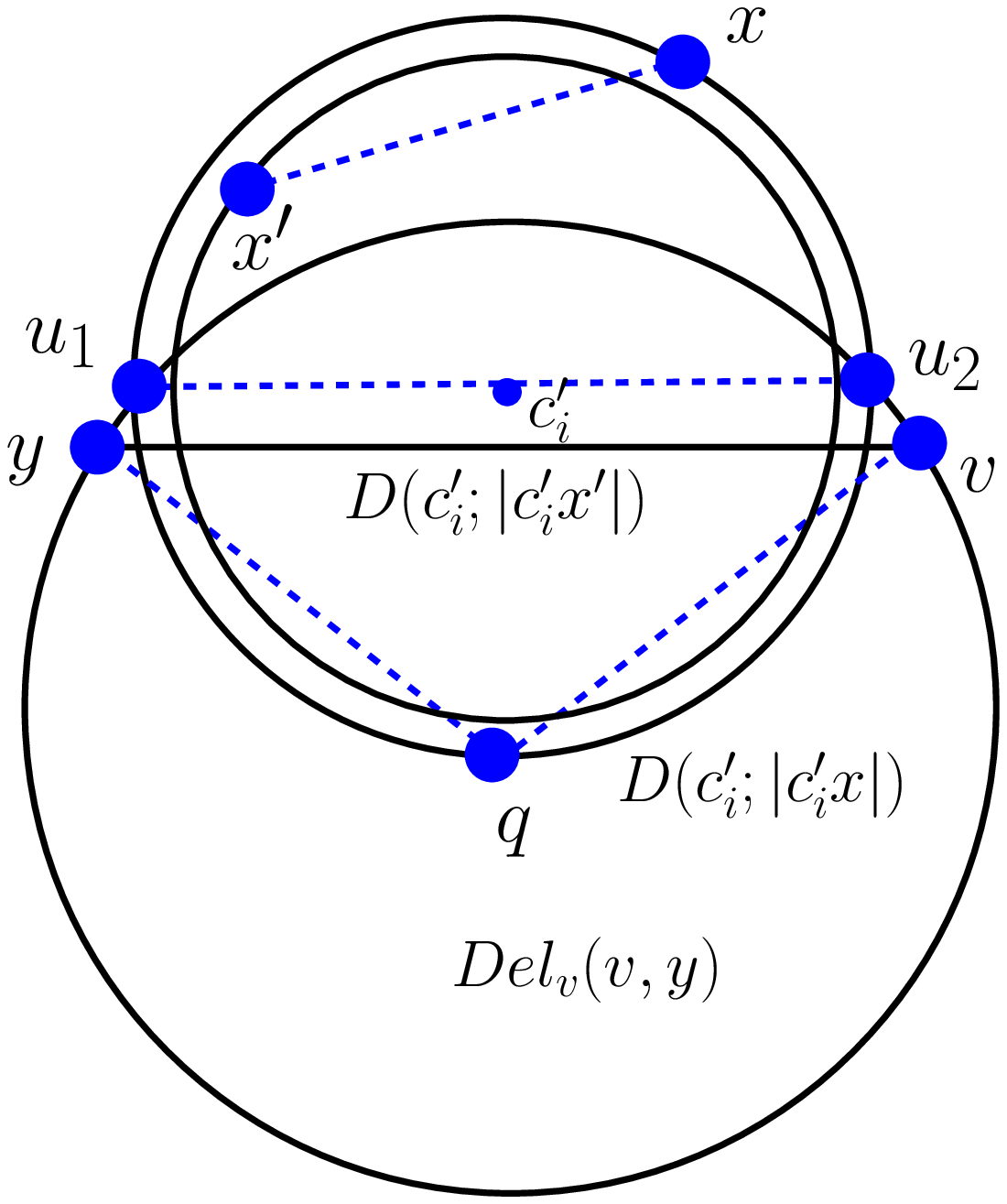}
    \end{minipage}}%
  \subfigure[$y=x$]{
    \label{fig08:b}
    \begin{minipage}[b]{0.5\textwidth}
      \centering
      \includegraphics[scale=0.50]{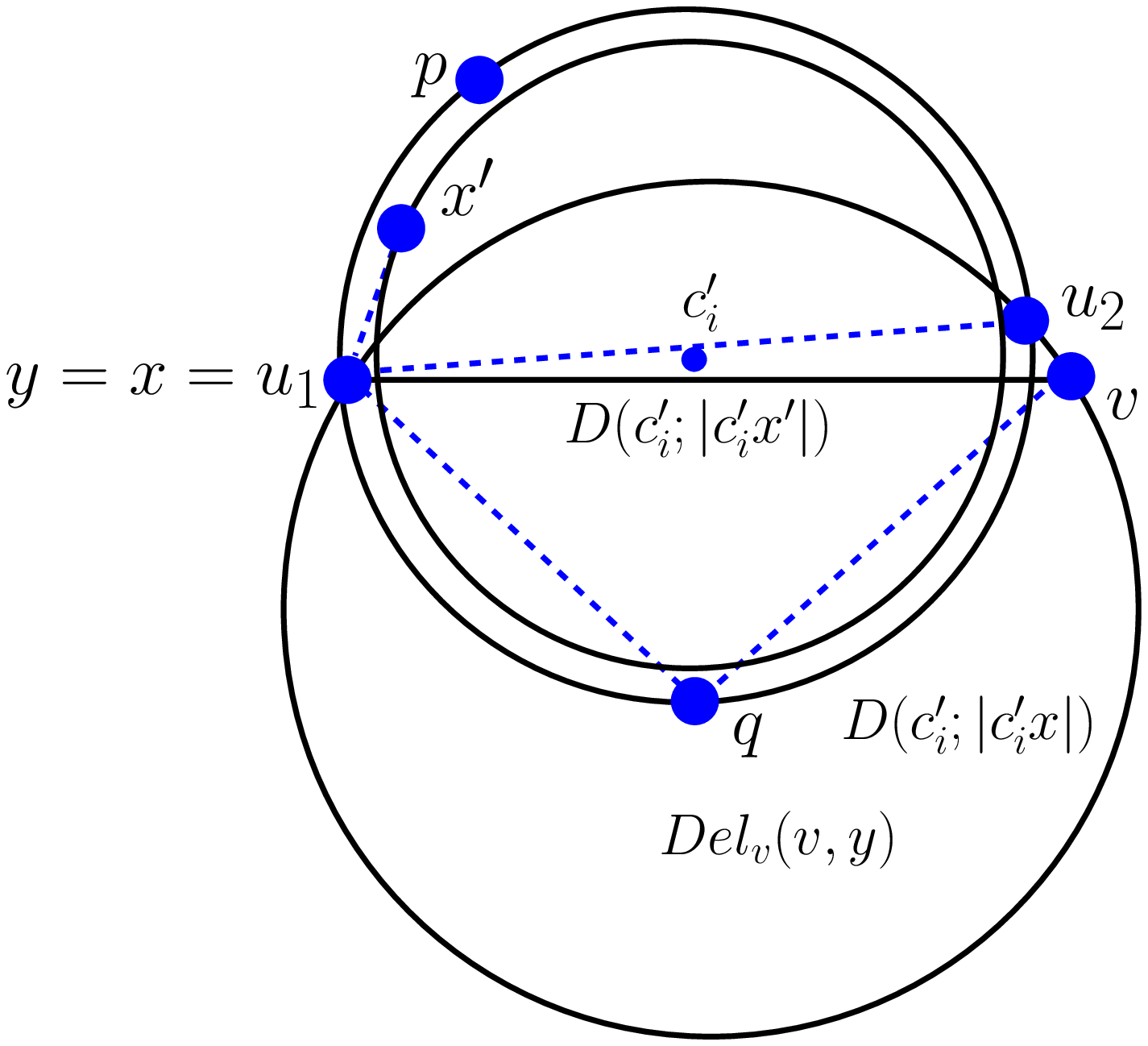}
    \end{minipage}}%
  \caption{\sl Illustrating the proof of Lemma \ref{lem09}.} 
  \label{fig08} 
\end{figure}

Since $|vq|>1$ and $|vy| \leq 1$, we have $\angle{yqv} < \pi/2$. 
Let $\widehat{yv}$ be the arc on $\partial \Del_v(v,y)$ with endpoints 
$y$ and $v$ that contains the north pole of $\partial \Del_v(v,y)$. 
Then $\widehat{yv}$ is a minor arc. 

Let $u_1$ and $u_2$ be the intersections between $\partial \Del_v(v,y)$
and $\partial D(x,p,q)$, where $u_1$ is to the left of $u_2$. 
Then both $u_1$ and $u_2$ are contained in $\widehat{yv}$ and, 
therefore, by Lemma~\ref{lem03}, $|u_1 u_2| \leq 1$. 

Let $\widehat{u_1 u_2}$ be the arc on $\partial D(x,p,q)$ with 
endpoints $u_1$ and $u_2$ that contains the north pole of 
$\partial D(x,p,q)$. Since $\angle{u_1 q u_2} \leq \angle{yqv} < \pi/2$, 
$\widehat{u_1 u_2}$ is a minor arc. 

Recall that $x \not\in \Int(\Del_v(v,y))$. Also, if 
$x \in \partial \Del_v(v,y)$, then $x=y$. It follows that $x$ is 
not below the line through $u_1$ and $u_2$. By a similar argument, 
$x'$ is not below this line. Since both $x$ and $x'$ are in 
$D(x,p,q)$ and since $\widehat{u_1 u_2}$ is a minor arc, it 
follows from Lemma~\ref{lem03} that $|xx'| \leq 1$, which is a 
contradiction. 

Thus, we have shown that $|c'_i x'| = |c'_i x|$. Recall that $p$ is 
the point that is computed in line~13 of algorithm $\aPLDG(v)$. 
Consider the point $p'$ that is computed in line~14 of algorithm 
$\aPLDG'(v)$. Since $D(c'_i;|c'_i x|) = D(c'_i;|c'_i x'|)$, we have 
$\{x,p\} = \{x',p'\}$. In other words, algorithm $\aPLDG'(v)$ knows the 
points $x$ and $p$, but does not know which of them is $x$ and which of 
them is $p$. 

Since lines 15--19 of algorithm $\aPLDG(v)$ are symmetric in $x$ and 
$p$, and since lines 16--22 of algorithm $\aPLDG'(v)$ are symmetric in 
$x'$ and $p'$, it follows that the behaviors of $\aPLDG(v)$ and 
$\aPLDG'(v)$ with respect to the edge $(v,y)$ are identical. 
Therefore, algorithm $\aPLDG'(v)$ removes the edge $(v,y)$ from 
$E'(v)$. 
\end{proof} 

\begin{lemma}    \label{lem09b}  
       Let $v$ be an element of $V$ and let $(v,y)$ be an edge of the 
       Delaunay triangulation $\LDT(v)$ of the set $N_v$. If algorithm 
       $\aPLDG'(v)$ removes $(v,y)$ from $E'(v)$, then algorithm 
       $\aPLDG(v)$ removes $(v,y)$ from $E(v)$. 
\end{lemma}
\begin{proof} 
Since algorithm $\aPLDG'(v)$ removes $(v,y)$ from $E'(v)$, there exist 
three pairwise distinct points $x$, $p$, and $q$ in $V$ such that 
\begin{enumerate} 
\item $\Delta(x,p,q)$ is a triangular face in $\LDT(x)$, 
\item algorithm $\aPLDG'(x)$ broadcasts the center $c'_i$ of the disk 
      $D(x,p,q) = D(c'_i; |c'_i x|)$, 
\item $v \neq x$, $|vx| \leq 1$, 
\item $v$ receives the center $c'_i$ (but does not know that it was 
      broadcast by $x$). 
\end{enumerate} 
Consider the point $x'$ that is computed in line~11 of algorithm 
$\aPLDG'(v)$. Thus, $x'$ is a point in $N_v \setminus \{v\}$ that is 
closest to $c'_i$. Since $x \in N_v \setminus \{v\}$, we have 
$|c'_i x'| \leq |c'_i x|$. In the rest of the proof, we will show that 
$|c'_i x'| = |c'_i x|$. As in the proof of Lemma~\ref{lem09}, this 
will imply that algorithm $\aPLDG(v)$ removes the edge $(v,y)$ from 
$E(v)$. 

The proof of the claim that $|c'_i x'| = |c'_i x|$ is by contradiction. 
Thus, we assume that $|c'_i x'| < |c'_i x|$. Then $x'$ is in the 
interior of the disk $D(x,p,q)$. Since $\Delta(x,p,q)$ is a triangular 
face in $\LDT(x)$, we have $|xx'|>1$. Since $|vx| \leq 1$, $v$ is not 
in the interior of $D(x,p,q)$. 

Consider the disk $D(c'_i; |c'_i x'|)$. It follows from line~13 of 
algorithm $\aPLDG'(v)$ that the boundary of this disk contains 
exactly two points of $N_v$; $x'$ is one of these two points, let 
$p'$ be the other one. Thus, $p'$ is the point that is computed in 
line~14 of algorithm $\aPLDG'(v)$. Since $y \in N_v \setminus \{v\}$, 
the point $y$ is not in the interior of $D(c'_i; |c'_i x'|)$. 

Consider the disk $\Del_v(v,y)$ that is computed in line~20 of 
algorithm $\aPLDG'(v)$. Then $N_v \cap \partial \Del_v(v,y) = \{v,y\}$ 
and $N_v \cap \Int(\Del_v(v,y)) = \emptyset$. Since $|vx'| \leq 1$ 
and $|vp'| \leq 1$, neither $x'$ nor $p'$ is in the interior of 
$\Del_v(v,y)$. 

Consider the point $z'$ on 
$\arc_i = ( \partial D(c'_i;|c'_i x'|) ) \setminus D(v;1)$ 
that is computed in line~18 of algorithm $\aPLDG'(v)$. It follows 
from line~21 that $z'$ is in the interior of $\Del_v(v,y)$ and 
$vy$ crosses at least one of $x'z'$ and $p'z'$. 

Since lines~11--22 of algorithm $\aPLDG'(v)$ are symmetric with 
respect to $x'$ and $p'$, we may assume without loss of generality that 
$vy$ crosses $x'z'$. Thus, $x' \not\in \{v,y\}$ and $x'$ and $z'$ 
are on opposite sides of the line through $v$ and $y$. 

We may assume without loss of generality that $vy$ is horizontal, 
$v$ is to the right of $y$, $x'$ is above the line through $v$ and 
$y$, and $z'$ is below this line. 

We claim that $y$ is in the interior of $D(x,p,q)$. The proof is by 
contradiction; thus, we assume that 
$y \not\in \Int(D(x,p,q))$. Observe that $z' \in \Int(D(x,p,q))$ 
and recall that $z' \in \Int(\Del_v(v,y))$. Since $|vz'|>1$ and 
$|vy| \leq 1$, we have $\angle{yz'v} < \pi/2$. Therefore, the upper arc 
on $\partial \Del_v(v,y)$ with endpoints $y$ and $v$ is a minor arc. 
Let $u_1$ and $u_2$ be the two intersection points between 
$\partial \Del_v(v,y)$ and $\partial D(x,p,q)$; refer to 
Figure~\ref{fig09}. It follows from Lemma~\ref{lem03} that 
$|u_1 u_2| \leq 1$. Since 
$\angle{u_1 z' u_2} \leq \angle{yz'v} < \pi/2$, the upper arc on 
$\partial D(x,p,q)$ with endpoints $u_1$ and $u_2$ is a minor arc. 
Since both $x$ and $x'$ are in $D(x,p,q)$ and on or above the line 
through $u_1$ and $u_2$, it follows, again from Lemma~\ref{lem03}, that 
$|xx'| \leq 1$, which is a contradiction. 

\begin{figure}
  \subfigure[$y \neq x$]{
    \label{fig09:a}
    \begin{minipage}[b]{0.5\textwidth}
      \centering
      \includegraphics[scale=0.50]{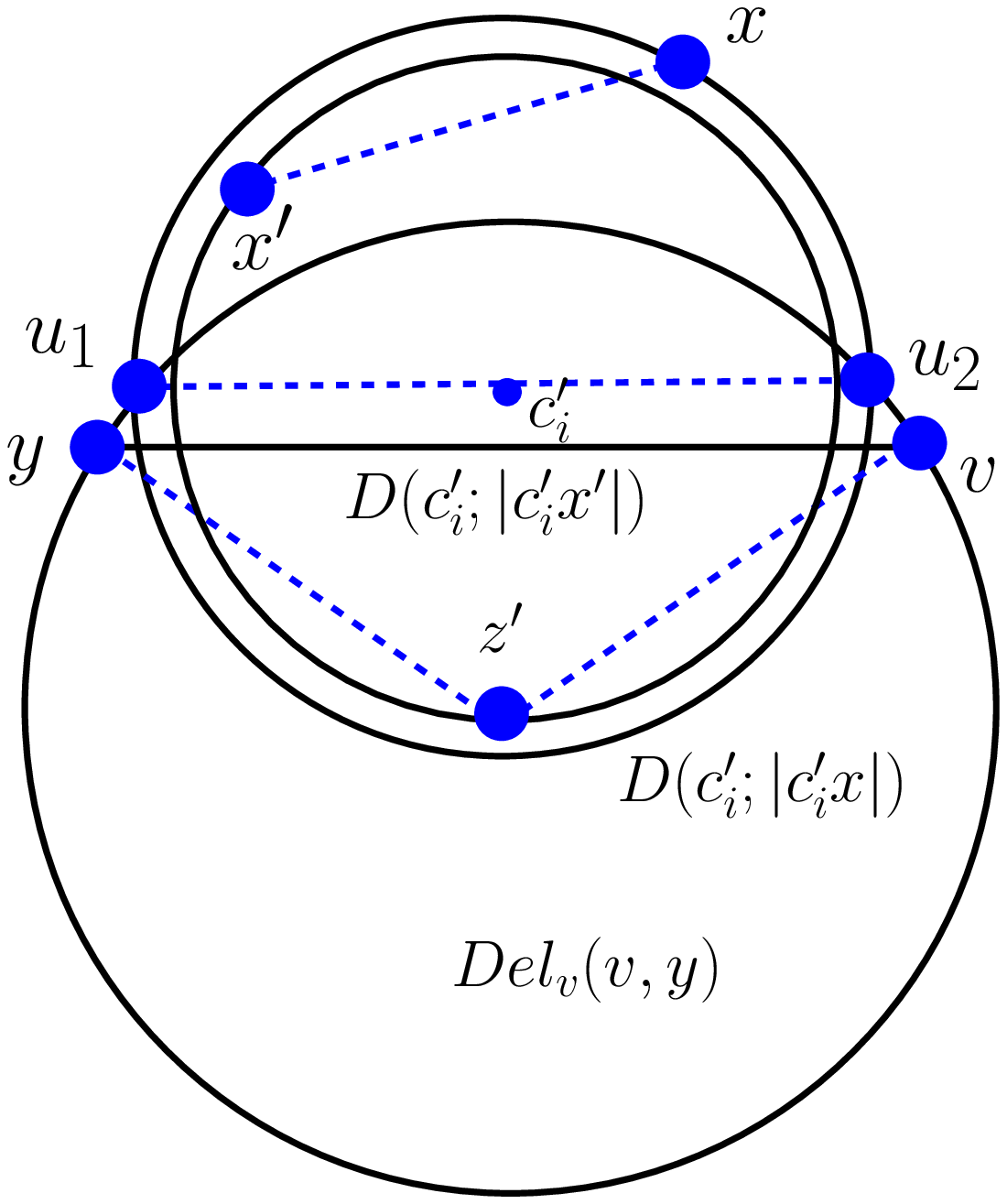}
    \end{minipage}}%
  \subfigure[$y=x$]{
    \label{fig09:b}
    \begin{minipage}[b]{0.5\textwidth}
      \centering
      \includegraphics[scale=0.50]{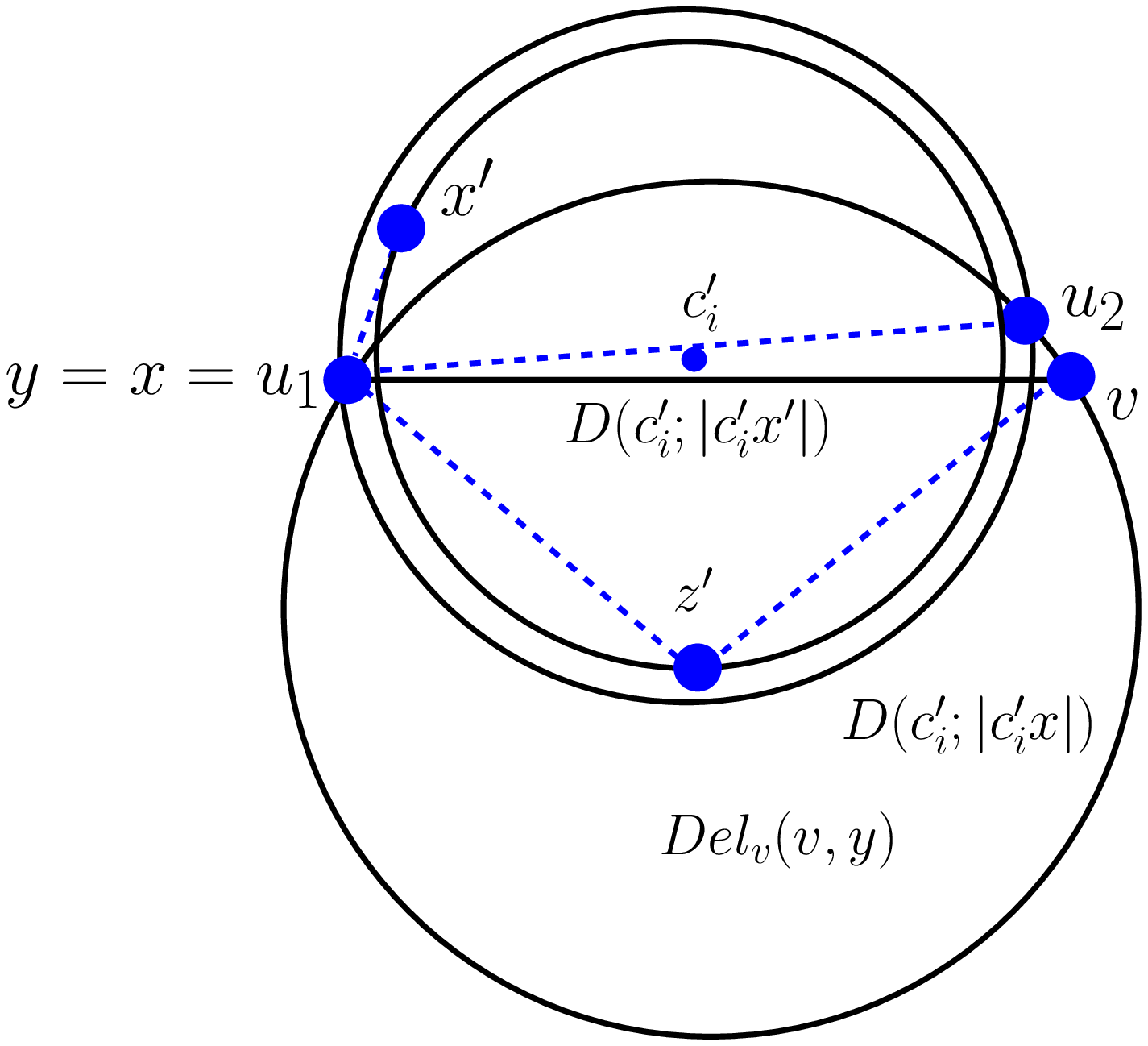}
    \end{minipage}}%

  \caption{\sl Illustrating the proof of Lemma \ref{lem09b}.} 
  \label{fig09}
\end{figure}

Thus, we have shown that $y \in \Int(D(x,p,q))$. Since $\Delta(x,p,q)$ 
is a triangular face in $\LDT(x)$, this implies that $|xy|>1$.  

We next claim that $x$ is below the line through $v$ and $y$. We prove 
this claim by contradiction. Thus, we assume that $x$ is above this 
line. Observe that $y$ is to the left of the vertical line through 
$c'_i$. We have seen above that $\angle{y z' v} < \pi/2$. Therefore, 
the arc on $\partial D(c'_i;|c'_i x'|)$ that is not below the line 
through $v$ and $y$ is a minor arc. It follows that $c'_i$ is below the 
line through $v$ and $y$. 

Consider the disk $D(x,p,q)$. We translate the center $c'_i$ of this 
disk horizontally to the right. During the translation, we change 
the disk so that $x$ stays on its boundary. We stop the translation 
as soon as one of $v$ and $y$ is on the boundary of the moving 
disk. Let $c^*$ be the center of the new disk $D^*$. Since $c^*$ is 
below the line through $v$ and $y$, the arc on $\partial D^*$ that is 
not below the line through $v$ and $y$ is a minor arc. First assume 
that $y$ is on the boundary of $D^*$. Then it follows from 
Lemma~\ref{lem03} that $|xy| \leq 1$, which is a contradiction.      
Thus, $y$ is in the interior of $D^*$ and $v$ is on the boundary 
of $D^*$. Then, the disk $D(y;|yv|)$ contains the point $x$ and, 
therefore, $|yx| \leq |yv| \leq 1$, which is also a contradiction. 

\begin{figure}
  \centering
  \includegraphics[scale=0.50]{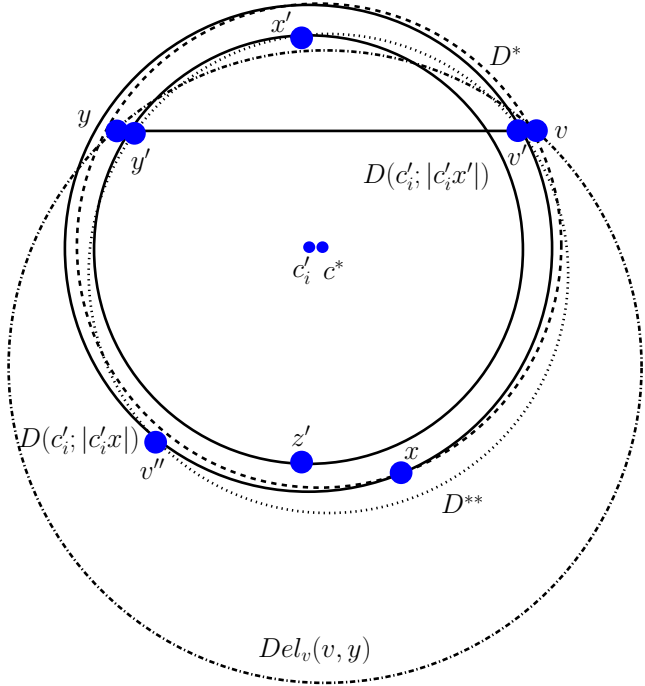}
  \caption{\sl An illustration of the proof of Lemma~\ref{lem09b}. 
               The point $y$ cannot be in the interior of 
               $D(c'_i;|c'_ix|)$.}  
  \label{fig10}
\end{figure}

We conclude that $x$ is below the line through $v$ and $y$. 
Let $y'$ be the leftmost intersection between $yv$ and 
$\partial D(c'_i;|c'_i x'|)$, and let $v'$ be the intersection between 
$yv$ and $\partial D(c'_i;|c'_i x|)$. 
We translate the center of the disk $D(c'_i;|c'_i x'|)$ along the line 
through $y'$ and $c'_i$, such that the center moves away from $y'$. 
During the translation, we change the disk so that $y'$ stays on its 
boundary. We stop the translation as soon as $v'$ is on the boundary of 
the moving disk; refer Figure~\ref{fig10}. Let $D^{**}$ be the 
resulting disk. Let $v'' \neq v'$ be the second intersection point 
between $\partial D(c'_i;|c'_i x|)$ and $\partial D^{**}$. 
Then $|y'v'|=|y'v''|$. Since $|yv'| \leq |yv| \leq 1$ and $|yx|>1$, 
the point $x$ is not in the disk $D(y';|y'v'|)$. Therefore, the 
point $x$ is on the clockwise arc on $D(c'_i;|c'_i x|)$ from $v'$ to 
$v''$. Observe that both $x$ and $x'$ are contained in $D^{**}$. 
It follows that any disk having $y'$ and $v'$ on its boundary contains 
at least one of $x$ and $x'$. This, in turn, implies that any disk 
having $y$ and $v$ on its boundary contains at least one of $x$ and 
$x'$. In particular, $\Del_v(v,y)$ contains at least one of $x$ and 
$x'$, which is a contradiction. This completes the proof of the lemma.  
\end{proof}

By Lemmas~\ref{lem09} and~\ref{lem09b}, algorithms $\aPLDG(v)$ and 
$\aPLDG'(v)$ compute the same graph. Therefore, the proof of 
Theorem~\ref{thm1} can be completed as in the proof of 
Lemma~\ref{lem10} and by observing that the sequence that is broadcast 
in line~8 of algorithm $\aPLDG'(v)$ contains at most $5$ points.

\section{Concluding Remarks}  \label{CONCL}
We have presented a $2$-local algorithm that constructs the Plane 
Localized Delaunay Graph $\PLDG(V)$ of any finite set $V$ of points in 
the plane. This graph is a plane and consistent 
$\frac{4 \pi \sqrt{3}}{9}$-spanner of the unit-disk graph $\UDG(V)$. 
Our algorithm makes only one communication round and each point of $V$ 
broadcasts at most $5$ messages. We leave as an open problem the 
question of whether a $2$-local algorithm exists in which each 
point broadcasts less than $5$ messages.   

In general, the maximum degree of any vertex in the graph $\PLDG(V)$ 
can be linear in the size of $V$. It is still open whether there is a 
communication-efficient localized algorithm that constructs a 
bounded-degree plane spanner of the unit-disk graph.

\bibliographystyle{plain}
\bibliography{LBDS}

\end{document}